\documentclass[11pt,amstex]{article}
\usepackage{amssymb,amsmath,cite}
\usepackage{latexsym}
\makeatletter\@addtoreset{equation}{section}\makeatother

\def\beq{\begin{eqnarray}}
\def\eeq{\end{eqnarray}}
\def\cL{{\cal L}}

\def\oneone{\rlap 1\mkern4mu{\rm l}}

\newtheorem{ex}{Example}[section]

\def\buildrel#1_#2^#3{\mathrel{\mathop{\kern 0pt#1}\limits_{#2}^{#3}}}

\newcommand{\ZZ}{\mathbb Z}
\newcommand{\Q}{\mathbb Q}

\newcommand{\p}{{\partial}{}}

\newcommand{\ls}{{\mathfrak{sl}(2,\mathbb{R})}{}}
\newcommand{\SL}{SL(2,\mathbb{R})}
\def \ra{{\rightarrow}}

\newcommand{\re}[1]{(\ref{#1})}

\newcommand{\be}{\begin{equation}}
\newcommand{\bea}{\begin{eqnarray}}
\newcommand{\ee}{\end{equation}}
\newcommand{\eea}{\end{eqnarray}}
\newcommand{\bes}{\begin{equation*}}
\newcommand{\ees}{\end{equation*}}
\newcommand{\beas}{\begin{eqnarray*}}
\newcommand{\eeas}{\end{eqnarray*}}
\newcommand{\pb}{\bar \partial}
\newcommand{\ap}{{\alpha_+}}
\newcommand{\Nb}{{\bar{N}}}
\newcommand{\mb}{{\bar{m}}}

\newcommand{\wads}{$\textit{WAdS}_3$}
\newcommand{\bb}{\mathbb}
\newcommand{\ph}{\phantom}
\newcommand{\bp}{\bar{\partial}}
\newcommand{\V}{{\cal V}}

\newcommand{\di}{\mathrm{d}}

\begin{document}


\begin{titlepage}
\begin{flushright}
July 16, 2010 \\
NSF-KITP-10-101
\end{flushright}
\vskip 1in

\begin{center}
{\Large{String Theory on Warped $\textit{AdS}_3$ and Virasoro Resonances}} 
\vskip 0.5in St\'ephane Detournay,$^{1}$ Dan Isra\"el,$^{2}$ Joshua M. Lapan,$^{1}$ and Mauricio Romo$^{3}$
\vskip 0.4in {\it $^{1}$ Kavli Institute for Theoretical Physics\\  University of California\\  Santa Barbara, CA 93106}
\vskip 0.2in {\it $^{2}$ GRECO, Institut d'Astrophysique de Paris \\ 
UMR 7095 CNRS -- Universit\'e Pierrre et Marie Curie\\
98bis Bd Arago, 75014 Paris, France}
\vskip 0.2in {\it $^{3}$ Department of Physics\\  University of California\\  Santa Barbara, CA 93106}
\end{center}
\vskip 0.5in

\begin{abstract}\noindent
We investigate aspects of holographic duals to time-like warped $AdS_3$ space-times --- which include G\"odel's universe --- in string theory.  Using worldsheet techniques similar to those that have been applied to $AdS_3$ backgrounds, we are able to identify space-time symmetry algebras that act on the dual boundary theory.  In particular, we always find at least one Virasoro algebra with computable central charge. Interestingly,  there exists a dense set of points in the moduli space of these models in which there is actually a second commuting Virasoro algebra, typically with different central charge than the first.  We analyze the supersymmetry of the backgrounds, finding related enhancements, and comment on possible interpretations of these results.  We also perform an asymptotic symmetry analysis at the level of supergravity, providing additional support for the worldsheet analysis.
	
\end{abstract}

\end{titlepage}

\section{Introduction and Summary}

The $\textit{AdS}/$CFT correspondence is a concrete and explicit realization of the holographic principle.  In the prototypical example of $\textit{AdS}_3$, it has been known for a long time that the asymptotic symmetry group consists of two Virasoro algebras with central charges $c=\bar{c}= \frac{3\ell}{2G}$ \cite{Brown:1986nw}.  That fact, combined with various other observations~\cite{Birmingham:2001pj, Carlip:2005zn}, suggests that pure gravity with negative cosmological constant and $\textit{AdS}_3$ boundary conditions is dual to a two-dimensional CFT, however the exact nature of that CFT remains a mystery and may not even exist~\cite{Maloney:2007ud,Gaberdiel:2007ve}.

Recently, tools developed in the context of $\textit{AdS}/$CFT have been exploited outside of their original high-energy/string theory realm, with potential groundbreaking applications to astrophysics, condensed matter, and collider physics~\cite{Guica:2008mu, Hartnoll:2009sz, Gubser:2009md}. 
A common feature of this trend is the extension of the holographic dictionary to non-$\textit{AdS}$ spaces, making it possible to relate a wider variety of geometries and dual field theories.  One such class of spaces, dubbed \emph{warped} $\textit{AdS}_3$, has attracted a lot of attention recently.  It consists of deformations of $\textit{AdS}_3$ that preserve a $\SL \times U(1) \subset \SL \times \SL$ subgroup of isometries.  Depending on the norm of the preserved $U(1)$ Killing vector, these spaces are referred to as time-like, space-like, or null $\textit{WAdS}_3$.  Though some have been known for quite a while ({\it e.g.}, in some region of parameter space, time-like $\textit{WAdS}_3$ is just G\"odel space-time~\cite{Godel:1949ga,Reboucas:1982hn,Rooman:1998xf}), the space-like representative has attracted a great deal of attention~\cite{Chen:2009cg,Gupta:2009zt,Gupta:2010ib,Chen:2009hg,Kao:2009fh, Miskovic:2009kr, Chakrabarti:2009ww, Anninos:2009zi,Blagojevic:2009ek, Compere:2008cv,Compere:2009zj} since the conjecture of Anninos {\it et al.} that topologically massive gravity (TMG) with negative cosmological constant and space-like $\textit{WAdS}_3$ boundary conditions is dual to a two-dimensional 
CFT~\cite{Anninos:2008fx}. The status of the conjecture is currently unclear since it has not been possible to give a clear geometric realization of the expected symmetries~\cite{Compere:2008cv,Compere:2009zj,Blagojevic:2009ek}. Understanding the holographic properties of space-like $\textit{WAdS}_3$ is also motivated by the fact that it appears as part of the near horizon extreme Kerr geometry (NHEK) and is believed to play a central role in the recently proposed Kerr/CFT correspondence~\cite{Guica:2008mu}.

In general, the question of identifying the field theory dual to a given gravitational background is a very intricate one. There is, however, a notable exception: when the given background can be embedded in string theory and realized as the near horizon geometry of some brane configuration. Consider an embedding of an $\textit{AdS}_3$ space-time in type IIB string theory on $\textit{AdS}_3 \times S^3 \times T^4$: in that case, in the duality frame with only Ramond-Ramond fluxes, the dual CFT can be identified as the worldvolume theory on the D1/D5 system which flows in the IR to an $\mathcal{N}=(4,4)$ SCFT which is in the moduli space of a symmetric product orbifold of  $T^4$~\cite{Aharony:1999ti, Seiberg:1999xz,David:2002wn}. In the duality frame which involves NS-NS fluxes only, the correspondence can be made even more precise on the gravity side by studying the worldsheet CFT, given by a $\SL \times SU(2) \times U(1)^4$ WZW model~\cite{Pakman:2009mi, Giribet:2007wp, Dabholkar:2007ey, Pakman:2007hn}. In this case, the (super-)Virasoro generators of the dual CFT actually act linearly on the Hilbert space of the worldsheet CFT, as was first shown by Giveon, Kutasov, and Seiberg in~\cite{Giveon:1998ns} and further developed in~\cite{deBoer:1998pp, Kutasov:1998zh,Kutasov:1999xu,Argurio:2000tb,Giveon:2003ku}. However, since the dual space-time theory is not at the orbifold point in its moduli space where it 
can be solved explicitly, a direct bulk/boundary correspondence can be made only for protected quantities, {\it e.g.} correlators of BPS states.

The aim of this paper is to explore some aspects of holography for $\textit{WAdS}_3$ spaces in the context string theory. These spaces can naturally be embedded in supergravity~\cite{Compere:2008cw,Anninos:2008qb, Levi:2009az,Orlando:2010ay} and, in fact, can be realized as exact string theory backgrounds as marginal deformations of the $\SL \times SU(2) \times U(1)^4$ WZW model~\cite{Israel:2003cx,Israel:2004vv,Detournay:2005fz, Compere:2008cw}.  Given the worldsheet CFT, we can explore the symmetries of the space-time theory and attempt to extract information about any holographically dual theory.  Although we don't expect to learn much about the exact nature of a theory dual to $\textit{WAdS}_3$ in TMG (it's not clear that TMG can be obtained as a consistent truncation of string theory), we can learn some quantitative features of the theory dual to $\textit{WAdS}_3$ in Einstein gravity (coupled to matter fields) which, in turn, could yield qualitative lessons for TMG.

All types of $\textit{WAdS}_3$ spaces --- time-like, space-like and null --- can be realized as exact string backgrounds through an elliptic, hyperbolic, or parabolic marginal deformation of the $\SL$ WZW model, respectively~\cite{Israel:2004vv}.  The hyperbolic and parabolic deformations are currently the most interesting from the space-time point of view: quotients of the hyperbolic deformation yield black hole solutions and
parts of the NHEK geometry, while both can yield gravity duals of non-relativistic systems~\cite{Balasubramanian:2008dm,Son:2008ye,D'Hoker:2010rz}. However, from the worldsheet point of view it is the elliptic deformation that is the most tractable because the spectrum of $\widehat \SL$ is unduly complicated in the hyperbolic or parabolic basis. 
This seemingly technical issue is one of the reasons why the description of the string spectrum on $\textit{AdS}_3$, including the subtle issue of the spectrally flowed sectors~\cite{Maldacena:2000hw,Maldacena:2000kv,Maldacena:2001km}, is better understood than it is for BTZ black holes (see however~\cite{Hemming:2001we,Rangamani:2007fz,Parsons:2009si}) even though the two spaces are related by discrete identifications. In particular, it is not even clear how the space-time Virasoro algebra arises from the worldsheet description of the BTZ black hole~\cite{Rangamani:2007fz} which is why, in this paper, we will exclusively study the elliptic (time-like) deformation of $\textit{AdS}_3$, hoping to one day return to tackle the null and space-like cases. 

Nevertheless, time-like warped {\it AdS}$_3$ is interesting on its own since this family of space-time metrics, obtained by 
Rebou\c cas and Tiomno as a one-parameter extension of the original G\"odel universe~\cite{Reboucas:1982hn}, are all homogeneous space-times with closed time-like curves (CTCs) through each point. It was shown in~\cite{Israel:2004vv} that the 
long string solutions in these space-times are the more relevant probes of this pathological space-time since they eventually wrap the 
CTCs (unlike the short strings which correspond to localized worldsheets). Thus, a holographic understanding of the worldvolume CFT living on a stack of long string would help us to understand how string theory deals with CTCs.

In section~\ref{WAdS3}, we review the embedding of time-like $\textit{WAdS}_3$ into string theory from both the supergravity and worldsheet perspectives.  We will see that only time-like stretched $\textit{WAdS}_3$ (in the terminology of~\cite{Anninos:2008fx}), corresponding to G\"odel spaces, can be supported by real matter fields. We also discuss the issue of the renormalization of the background fields, showing that this embedding of $\textit{WAdS}_3$ corresponds to an exact string background.

In section~\ref{Bosonic}, we address the main point of this paper for the bosonic string, namely the construction of a space-time symmetry algebra from the $\textit{WAdS}_3$ worldsheet theory. We start by reviewing the contruction of Giveon, Kutasov, and Seiberg for pure $\textit{AdS}_3$~\cite{Giveon:1998ns}. It was shown there that the two space-time Virasoro algebras have central charges $c_{\mathrm{st}} = \bar{c}_{\mathrm{st}} = 6 k |w|$, where $k$ is the level of the $\widehat{\SL}$ and $w$ is the spectral flow sector of the string, representing the number of times the string worldsheet wraps around the origin. This corresponds to the central charge of the worldvolume 
theory living on $|w|$ long strings near the boundary~\cite{Seiberg:1999xz}.

Turning to the analysis of the elliptic deformation of the $\SL\times U(1)$ WZW model, which corresponds to fibering an $S^1$ over a time-like $\textit{WAdS}_3$, we find that there always exists at least one space-time Virasoro algebra with central charge $\bar{c}_{\mathrm{st}}=6k|w|$ arising from the anti-holomorphic side of the theory.
From the holomorphic side, the only space-time charges that can generically be constructed are a global $U(1)^2$ that correspond to the zero-mode of a would-be space-time Virasoro algebra, $L_0^{\mathrm{st}}$, and the zero-mode of a would-be space-time $\widehat{U(1)}$, $J_0^{\mathrm{st}}$. The existence of both $L_0^{\mathrm{st}}$ and $\bar{L}_0^{\mathrm{st}}$ suggests that the dual theory is at least a two-dimensional Lorentz-invariant theory, if not conformal. However, we observe that when the $S^1$ radius $R$ and squashing parameter $\textsc{h}$ satisfy
\begin{equation}\label{SymmEnh}
\frac{2\textsc{h}R}{\sqrt{k}} = \frac{\mu}{\rho} \in \Q \, ,
\end{equation}
for $\mu$ and $\rho$ relatively prime positive integers, it becomes possible to construct a second Virasoro algebra with central charge
\begin{equation}\left\{ 
\begin{array}{lcc} \hat{c}_{\mathrm{st}}= 6k (\rho w + \mu \omega ) \, , &\quad  w>0\, ,  \\
 \hat{c}_{\mathrm{st}}= -6k (\rho w + \mu \omega )\, ,  &\quad  w<0 \, , 
\end{array} \right.
\end{equation}
where $\omega$ represents the winding number around around the fibered $S^1$.  Thus, for a given $\textsc{h}$ and $k$, we can tune the $S^1$ radius $R$ so that, in appropriate units, for irrational values there is only one space-time Virasoro algebra, while for rational values there are two copies as in a two-dimensionsal CFT.

In section~\ref{superstring}, we extend the bosonic construction to heterotic and type II superstrings on $\textit{WAdS}_3 \times S^1 \times S^3 \times T^4$.  In the type II case, we find that space-time supersymmetry is broken in half by the deformation \emph{except} when $\frac{2\textsc{h}R}{\sqrt{k}} \in \mathbb{Z}$, in which case all supersymmetries are preserved.  In the heterotic case, because of the asymmetry of the worldsheet theory, we can choose to deform the anti-holomorphic $\SL$ and preserve all space-time supersymmetries, or we can deform the holomorphic (supersymmetric) $\SL$ and generically break all space-time supersymmetries, except when $\frac{2\textsc{h}R}{\sqrt{k}}$ is integer. 

In section~\ref{sec:ASG}, we perform an asymptotic symmetry analysis of the ten-dimensional background, much in the spirit of Brown and Henneaux \cite{Brown:1986nw}. We find that when \re{SymmEnh} is satisfied and  $\rho = 1$, the $\ls \oplus u(1) \oplus u(1)$ isometry algebra is enhanced to $\ls \oplus \ls \oplus u(1)$.  In this case, $\ls \oplus \ls$ naturally extends to two copies of the Virasoro algebra with central charges
\begin{equation}
c_{\mathrm{st}} =  \bar{c}_{\mathrm{st}}= \frac{3 \ell \sqrt{1 + 2\textsc{h}^2}}{2 G_3},
\end{equation}
where $G_3$ is the three-dimensional Newton's constant and $\ell^2 = k$. When $\rho \ne 1$ but (\ref{SymmEnh}) still holds, the global isometry is restricted to  $\ls \oplus u(1) \oplus u(1)$ but there is still a natural extension to two Virasoro algebras where $c_{\mathrm{st}} = \rho \bar{c}_{\mathrm{st}} = \rho\frac{3 \ell \sqrt{1 + 2\textsc{h}^2}}{2 G_3}$, coinciding with the worldsheet interpretation when there is no winding around the fibered $S^1$.  Then we check the supersymmetries preserved by the backgrounds and find the same results as in section \ref{superstring}.

Finally in section \ref{sec:discussion}, we discuss possible interpretations of our results as well as open questions.  Further background material and complimentary results are collected in the
appendices.


\section{Warped $\textit{AdS}_{3}$ as an Exact String Background}\label{WAdS3}

In this section we review an embedding of time-like warped $\textit{AdS}_{3}$ in string theory.  This space-time is a solution of three-dimensional Einstein gravity with a Maxwell-Chern-Simons $U(1)$ gauge field and scalars, obtained as a consistent truncation of ten-dimensional supergravity compactified on $S^{1}\times S^{3}\times T^{3}$.  Since the ten-dimensional solution only involves non-trivial NS-NS fields, the analysis applies equally to type II or heterotic string theories.  After the supergravity analysis we will demonstrate that this background corresponds to an exact string solution by realizing the worldsheet sigma model as a deformed WZW model, the renormalization properties of which are well known \cite{Tseytlin:1992ri,Tseytlin:1993my,Forste:2003km}. 

\subsection{Supergravity: 3d Kaluza-Klein Reduction and Solution}
\label{sec:kk-reduction}

Let us begin with the action for the common bosonic sector of type IIA/B and heterotic supergravities in Einstein frame,
\begin{equation}
\label{10dtheory}
S = \frac{1}{2\kappa_{10}^2} \int \di^{10}x \, \sqrt{-g} \bigg[ R_{10} - \frac{1}{2} \big( \p \Phi \big)^2 - \frac{1}{12} e^{-\Phi} H_{MNP} H^{MNP} \bigg].
\end{equation}
Then we compactify on $S^3 \times T^3 \times S^1$ while retaining the KK-gauge field from the $S^1$, taking the ten-dimensional background to be\footnote{The truncation of the KK-reduced theory to massless modes is known to be consistent for $S^3$ \cite{Cvetic:2000dm}.}
\begin{subequations}
\label{10dBG}
\begin{eqnarray}
\di s_{10}^{2}&=&  e^{-\frac{3Y}{2}} \left[ e^X \di s_3^2 + e^{-X} (\di \varphi+A)^2 \right]+e^{Y}L_S^{\, 2} \di s^2 (S^{3})+\di s^2 (T^{3}) \, , \\
H &=&  h_S L_S^{\, 3} \textrm{Vol}(S^3) + \hat{H} + \hat{F} \wedge (\di \varphi+ A)  \, ,
\end{eqnarray}
\end{subequations}
where $\hat{H} \equiv d\hat{B} - \hat{F} \wedge A$, $L_S$ is the radius of $S^{3}$, $h_S$ is a constant, and $\di s^2 (S^{3})$ and $\textrm{Vol}(S^3)$ denote the metric and volume form on the unit three-sphere, respectively.  
This reduces to
\begin{multline}
S_{3d} = \frac{4\pi^3 L_S^{\, 3} R V_{T^3}}{2\kappa_{10}^2} \int \di^3 x\, \sqrt{-g_3} \bigg[ R_3 -\frac{1}{4}e^{-2X}F^{2} -
\frac{1}{4}e^{3Y-\Phi} \hat{F}^{2}
-\frac{1}{12} e^{-\Phi} \left(\hat{H}^{2}e^{3Y-2X}+6h_{s}^{2}e^{X-\frac{9Y}{2}}\right)
  \\  +\mathcal{L}_{kin}(Y,X,\Phi)+\frac{6}{L_S^{\, 2}}\, e^{X-\frac{5Y}{2}}\bigg] \, .
\end{multline}

The equation of motion for $\hat{B}$ requires $e^{-\Phi-2X+3Y} \star_3 \hat{H} = h_3 = \mathit{constant}$, so we can simply integrate out $\hat{H}$.  We can further truncate (consistently) to a theory where $e^{X+\frac{3}{2}Y-\frac{1}{2}\Phi} = 2$ and $\hat{A} = \frac{1}{2} A$, arriving at the action
\bea
\label{3daction}
S_{3d} &=& \frac{1}{2\kappa_{3}^2} \int \di^3 x \, \sqrt{-g_3} \bigg[ R_3 - \frac{1}{8} e^{3 Y-\Phi} F^2 + \mathcal{L}_{kin}(\Phi,Y) - 2 h_3^2 e^{2\Phi-6 Y} + \frac{12}{L_S^2} e^{-4 Y+\frac{1}{2}\Phi}   \\
&& \qquad\qquad \qquad \qquad - h_S^2 e^{-6 Y - \frac{1}{2} \Phi} \bigg]  -  \frac{h_3}{4\kappa_3^2}  \int A \wedge F \, .
\eea
This consistent truncation admits the \wads~solution
\begin{subequations}
\label{SugraSolWads}
\begin{eqnarray}
\di s_3^2  &=& \frac{k}{8} e^{3Y-\Phi}  \left[ \di r^2 + \sinh^2r\,
\di \alpha^2 - (1+2 \textsc{h}^2) (\di \beta + \cosh r\, \di \alpha)^2\right] \, , \\ 
A &=& \sqrt{k} \,\textsc{h}  (\di \beta + \cosh r \di \alpha)\, , \\ 
e^{2 Y} &=& \frac{h_S^2}{4} e^{-\Phi} \, ,
\end{eqnarray}
\end{subequations}
where $\Phi$ is a constant, $e^{Y}$ and $h_S$ have been rescaled to set $L_S=1$ (in units of $\alpha'$), and
\begin{subequations}
\bea
k &=&\frac{h_S}{2}  \, , \\
\textsc{h}^2 &=& \frac{2e^{\frac{5}{2}\Phi} h_3^2-h_S^2}{2 h_S^2} \, .
\eea
\end{subequations}
As measured in string-frame, the ten-dimensional metric is the product of a $T^4$ at arbitrary volume with an $S^3$ at radius $\sqrt{k}$ and an $S^1$ bundle over \wads~at radius $\sqrt{k}$.

Heterotic supergravity can also be dimensionally reduced on $S^3 \times T^4$ to similarly obtain Einstein-Maxwell-CS theory with scalars, the main difference being that the gauge field arises from a ten-dimensional gauge field instead of from a Kaluza-Klein circle.  The truncated three-dimensional theory will also admit \wads~as a solution.

\subsection{Worldsheet CFT Description: Deformed WZW Model}

We now review how this solution is obtained as a solvable worldsheet conformal field theory following~\cite{Israel:2003cx}. 
Let's begin with a bosonic $\SL$ WZW model corresponding to string propagation in an $\textit{AdS}_3$ space-time. 
Using the Euler angles parameterization, $g=e^{i\sigma_3 \alpha/2} e^{\sigma_1 r/2}e^{i\sigma_3 \beta/2}$, the worldsheet action reads
\begin{eqnarray}
S_{\SL_k}(r,\alpha,\beta) = \frac{k}{8 \pi} \int \di^2 z\, \left[ \p r \pb r -\p \beta \pb \beta - \p \alpha \pb \alpha - 2 \cosh r \p \alpha \pb \beta \right] \, .
\end{eqnarray}
The background fields are, thus,\footnote{$\alpha'=1$ in this section.}
\begin{subequations}
\begin{eqnarray}
\di s^2 &=& \frac{k}{4} (\di r^2 + \sinh^2r \di \alpha^2 - (\di \beta + \cosh r \, \di \alpha)^2)\, , \\
B &=&  \frac{k}{4} \cosh r\, \di \beta \wedge \di \alpha \,.
\end{eqnarray}
\end{subequations}
This is $\textit{AdS}_3$ in global coordinates, taking its more familiar form if we write $\alpha = t-\phi,~\beta=t+\phi,$ and $r=2\varrho$.

Next, we add a free, compact boson $\varphi$ with radius $R$ and then deform by an exactly marginal, asymmetric, current-current deformation to obtain a $W\!AdS_{3}$ background:
\begin{eqnarray}\label{DefAs}
S = S_{SL(2)}(r,\alpha,\beta) +  \frac{1}{4 \pi} \int \di^2 z \, \p \varphi \pb \varphi - \frac{2 \textsc{h}}{ \pi \sqrt{k}} \int \di^2 z \,  J^3 (z) i\pb \varphi  \, ,
\end{eqnarray}
where $J^3 =  \frac{ik}{4}(\p \beta + \cosh r \p\alpha)$ is the holomorphic elliptic/time-like current of $\SL$. This deformation corresponds to turning on $B_{\mu\varphi}=G_{\mu\varphi} = G_{\phi\phi}A_\mu$, where $G_{\varphi\varphi}=\frac{1}{2}$, $\mu=r,\alpha,\beta,$ and
\begin{equation}
A = \sqrt{k}\, \textsc{h} (\di \beta + \cosh r\,  \di \alpha) \, .
\end{equation}

A Kaluza-Klein reduction down to three dimensions yields the background\footnote{In heterotic string theory, one can take $\varphi(\bar z)$ to be an antiholomorphic chiral boson corresponding to heterotic gauge-bundle degrees of freedom. In this case, the analysis is more subtle but eventually leads to the same three-dimensional background fields as this Kaluza-Klein reduction.}
\begin{subequations}
\begin{eqnarray}\label{WZWWadsSol}
\di s^2 &=&(G_{\mu \nu} -  G_{\varphi \varphi} A_\mu A_\nu)\di x^\mu \di x^\nu \nonumber \\ &=&  \frac{k}{4} \left[ \di r^2 + \sinh^2r\,
\di \alpha^2 - (1+2 \textsc{h}^2) (\di \beta + \cosh r\, \di \alpha)^2\right] \, , \\
B &=&  \frac{k}{4} \cosh r \di \beta \wedge \di \alpha \, , \\
A &=& \sqrt{k}\, \textsc{h}  (\di \beta + \cosh r \di \alpha) \, .
\end{eqnarray}
\end{subequations}
This solution is exactly the same as the solution in~(\ref{SugraSolWads}). 
Of course in string theory models like $\textit{AdS}_3\times S^3 \times T^4$, the WZW level $k$ is quantized to integer values.

\subsection{Renormalization of the Background Fields}
For a generic WZW model, it is known that the background fields extracted from the classical action are actually exact to all orders in $\alpha'$, modulo a shift of the level  $k \ra k + c_g$, where $c_g$ is the dual coxeter number.

In the same way, one might wonder how the background fields are renormalized for the  $WAdS_3$ solution.  For this, we notice that~\re{DefAs} can be rewritten as~\cite{Israel:2004vv}
\begin{equation}\label{DefAs2}
S = S_{SL(2)}\left(r,\alpha,\beta - \frac{2\textsc{h}}{\sqrt{k}}  \varphi\right) + \frac{1}{4 \pi} (1+2\textsc{h}^2)  \int \di^2 z \, \p  \varphi \pb \varphi  +  \frac{\textsc{h}\sqrt{k}}{4\pi} \int\di^2 z \left( \p\beta\pb\varphi - \pb\beta\p\varphi \right)  \, ,
\end{equation}
where the last term is a total derivative whose {\it raison d'etre} is to contribute a phase to the partition function in winding sectors.
Being the formal sum of two WZW models (modulo the total derivative term), the renormalization in the effective action can be computed as usual, resulting in a shift $k \ra k -2$ in the $\widehat{\SL}$ level. This leads to a renormalization of the four-dimensional fields which, in turn, yields the three-dimensional background:
\begin{subequations}
\begin{eqnarray}
\di s^2 &=& \frac{k -2}{4} \left[\di r^2 + \sinh^2r\, \di \alpha^2 - \left(1+2 \frac{(k-2) \textsc{h}^2}{k+4 \textsc{h}^2}\right) (\di \beta + \cosh r\,  \di \alpha)^2\right] \, , \\
B &=&  \frac{k-2}{4} \cosh r\, \di \beta \wedge \di \alpha \, , \\
A &=& (k-2) \sqrt{k} \frac{\textsc{h}}{k+4\textsc{h}^2}  (\di\beta + \cosh r \, \di \alpha)\, ,
\end{eqnarray}
\end{subequations}
which coincides with the classical expressions in the large $k$ limit, as it must. As promised, the renormalization corresponds only to a redefinition of the parameters, hence, we can treat the solution~(\ref{SugraSolWads}), or
equivalently~(\ref{WZWWadsSol}), as an exact perturbative string background.


\section{Space-time Symmetry Algebra I:~~Bosonic Strings}\label{Bosonic}

We now turn to the main part of this work and investigate the space-time symmetries associated with these $\textit{WAdS}_3$ string backgrounds.  For simplicity, we start with bosonic string theory on $\textit{WAdS}_3 \times U(1) \times \mathcal{M}$ where $\mathcal{M}$ is an arbitrary unitary CFT with $c=22-\frac{6}{k-2}$.

In~\cite{Giveon:1998ns}, the authors constructed the space-time Virasoro algebra of $\textit{AdS}_3$ in the free-field limit of the worldsheet CFT.
This free-field limit corresponds to a long string near the boundary of $\textit{AdS}_3$ that winds $w$ times around the origin.  As such, the central charge `felt' by the string depends on its winding as the string effectively forms a $w$-sheeted cover of the boundary~\cite{Apikian:1996yg}.  In section~\ref{sec:gksreview},
we briefly review the calculation of~\cite{Giveon:1998ns} in coordinates that will suit our purposes when confronting the deformed model, studied in section~\ref{sec:bosonicdeformed}.
In appendix~\ref{app:bosonicbrst}, we construct the charges using solely $SL(2,\bb{R})$ primaries without reference to free fields.


\subsection{Lightning Review of the Giveon-Kutasov-Seiberg Construction}
\label{sec:gksreview}
We review here the construction of the space-time Virasoro algebras in $\textit{AdS}_3$ using a free-field description of the bosonic string in an $\textit{AdS}_3 \times U(1) \times \mathcal{M}$ background \cite{Giveon:1998ns}.
We will consider a single copy of the algebra that is built out of holomorphic worldsheet operators, remembering that there is a second copy built from analogous antiholomorphic operators.

We start with the holomorphic $\widehat{SL(2,\bb{R})}$ currents in the elliptic/time-like basis which satisfy the algebra
\begin{equation}
J^3(z) J^3(0) \sim -\tfrac{k}{2z^2} \, ,  \qquad  J^3(z) J^\pm(0) \sim \pm\tfrac{1}{z} J^\pm(0) \, ,  \qquad  J^+(z) J^-(0) \sim \tfrac{k}{z^2} - \tfrac{2}{z} J^3(0) \, .
\end{equation}
We can express the elliptic Cartan current as $J^3= i\sqrt{\frac{k}{2}}\, \p T$ by introducing a  free, time-like, chiral boson $T(z)$ with OPE $T(z)T(0)\sim \ln z$.\footnote{Throughout this section and the following ones, we work in units where $\alpha'=2$. By an abuse of notation, the antiholomorphic counterpart of a chiral boson $\varphi(z)$ will be denoted by $\bar{\varphi} (\bar z)$ while the associated non-chiral boson will be denoted by $\varphi (z,\bar z)$.} This  allows us to decompose the remaining currents, $J^\pm (z)$, in terms of $SL(2,\mathbb{R})/U(1)$
parafermions as (see appendix~\ref{app:parafermions} for background on parafermions):
\begin{equation}
J^{+} (z) = i\sqrt{k}\,\psi_1 \,e^{- i \sqrt{\frac{2}{k}} T}(z) \, , \qquad J^{-} (z) = -i\sqrt{k}\,\psi_1^{\dag} \, e^{ i \sqrt{\frac{2}{k}} T} (z) \, .
\end{equation}

We now use the free-field representation of non-compact parafermions, valid near the boundary of the target space (see (\ref{freeparafrep}) in appendix~\ref{app:parafermions}), to study the long string sectors of the 
worldsheet theory. In terms of the free bosons, $T$ and $\chi$, and the linear dilaton, $s (z,\bar z)$, an $\widehat{SL(2,\bb{R})} \times \widehat{SL(2,\bb{R})}$ primary of spin $j$ is represented as
\begin{equation}
\label{freefieldprimaries}
\V_{j,m,\bar m}^w (z,\bar z) =  e^{\frac{2}{\alpha_+} j\, s(z,\bar z)} \,   e^{i\sqrt{\frac{2}{k}}\left( m \chi (z) + \bar m \bar{\chi} (\bar z) \right) } e^{-i\sqrt{\frac{2}{k}}\left( (m+\frac{kw}{2} ) T (z) + (\bar m +\frac{k w}{2} )\bar{T} (\bar z) \right) } \, ,
\end{equation}
where $\alpha_+ \equiv \sqrt{2k - 4}$. Representations are labeled by their spectral flow, $w \in \mathbb{Z}$, which corresponds to an outer automorphism of the left and right affine algebras.\footnote{On the single cover of the $\SL$, we could have different spectral flow parameters, $w$ and $\bar{w}$, for left- and right-moving sectors, provided that level-matching is satisfied.  However, working on $\textit{AdS}_3$, the universal cover of $\SL$, forces $w=\bar{w}$.} The left and right conformal dimensions of these primaries are
\begin{subequations}
\label{confdims}
\begin{eqnarray}
\Delta &=& -\frac{j(j-1)}{k-2} - wm -\frac{k w^2}{4} \, ,  \\
\bar{\Delta} &=& -\frac{j(j-1)}{k-2} - w\bar{m} -\frac{k w^2}{4} \, .
\end{eqnarray}
\end{subequations}

In order to realize the space-time Virasoro algebra as charges acting on the worldsheet Hilbert space, we need to identify holomorphic $(1,0)$ currents which are obtainable as Ka\v c-Moody descendents of the holomorphic $\widehat{SL(2,\mathbb{R})}$  primaries, $\V^{0}_{0,m,0}$. As this operator has $\bar m=0$, consistency of the theory requires that $m=n \in \mathbb{Z}$.\footnote{This constraint comes from enforcing level matching $\Delta - \bar{\Delta}\in\bb{Z}$ under spectral flow by $J^3$ and $\bar{J}^3$.  It also corresponds to the quantization of angular momentum in global $\textit{AdS}_3$ \cite{Maldacena:2000hw}.}  In the free-field formulation, the most general charge that can be obtained is
\begin{equation}
L^{\mathrm{st}}_n \equiv \oint \frac{\di z}{2i\pi} \bigg[ b_n^+ (\p \chi + \p T)
+ b_n^- (\p \chi - \p T) -i b^s_n \p s \bigg] e^{i\sqrt{\frac{2}{k}}\, n (\chi-T)} \, .
\end{equation}
The condition for the operator to be a Virasoro primary and, hence, BRST invariant is
\begin{equation}
n \ap  \, b_n^+ + \sqrt{\frac{k}{2}}\, b_n^s = 0 \, .
\end{equation}
The $\p (\chi-T)$ term is the integral of a total derivative --- equivalently, a BRST exact term --- for any $n\neq 0$
and can be chosen freely. We make the gauge choice $b_n^-=0$ for $n \neq 0$.

Computing the commutator $[L^{\mathrm{st}}_n, L^{\mathrm{st}}_{n'}]$ and demanding it form a Virasoro algebra yields the constraints
\begin{equation}
b_0^- = \frac{i}{2}\sqrt{\frac{k}{2}}\quad ,  \qquad  2i \sqrt{\frac{2}{k}} b_m^+ b_{n '}^+ = b_{n+n'}^+ \, ,
\end{equation}
the latter of which can be solved by setting $b_n^+ = -\tfrac{i}{2} \sqrt{\frac{k}{2}}$.  Thus, we have the worldsheet charges
\be
L^{\mathrm{st}}_n = \frac{k\hat{p}}{4} \delta_{n,0} -  \oint \frac{\di z}{2i\pi } \bigg[ \frac{1}{2}\sqrt{\frac{k}{2}}\, i(\p \chi+ \p T)   -  n\, \sqrt{\frac{k-2}{2}}   \p s \bigg] e^{i\sqrt{\frac{2}{k}}\, n (\chi-T)}
\ee
which satisfy a Virasoro algebra with central charge $c_{\mathrm{st}}=6k\hat{p}$, where $\hat{p} \equiv i \sqrt{\frac{2}{k}} \oint \frac{\di z}{2i\pi } (\p \chi-\p T)$.  The value of $\hat{p}$ depends on the spectral flow sector of the string, which can easily be seen by considering the operator inside a correlation function
\begin{equation}
\label{eqn:central-charge-winding}
\Big\langle  i \sqrt{\frac{2}{k}} \oint \frac{\di z}{2i\pi } (\p \chi-\p T)(z) \V^w_{j,m,\bar{m}}(0) \Big\rangle  = w \Big\langle  \V^w_{j,m,\bar{m}}(0) \Big\rangle\, .
\end{equation}
In fact, the value $w$ has the physical interpretation of the number of times the worldsheet of the long string wraps around the origin.  The central charge is then\footnote{The relation between the space-time and worldsheet time coordinates for a long string in the $w$-sector 
is $t=w \tau$. When $w>0$, the origin of the planar worldsheet of the long string is the infinite past, so
that is where we should insert the twist operator that defines our long string state.  On the other hand, when $w<0$,
the infinite space-time past is at future infinity on the planar worldsheet, implying that we should insert the twist operator there.  Relative
to future infinity, the contour used in computing the central charge (\ref{eqn:central-charge-winding}) is oriented
in a \emph{clockwise} direction, thus contributing a second minus sign (in addition to $w=-|w|$) and ensuring the central charge is positive.}
\begin{equation}
\label{centralcharge}
c_{\textrm{st}} = 6k|w| \, .
\end{equation}
Of course, the entire discussion above can be repeated for the antiholomorphic sector of the theory, leading to a commuting space-time Virasoro algebra $\bar{L}^{\mathrm{st}}_m$ with central charge $\bar{c}_{\mathrm{st}} = 6k|w|$.  As we said, this computation is for a given spectral flow $w$, however, the spectral flow is not conserved by the interactions (which take place far from the free-field limit region $s\to -\infty$) and, thus, does not correspond to a `superselection' sector of the theory.

As said in the introduction, having a brane construction of the worldsheet theory leads to interesting insights.  Consider the case of $\textit{AdS}_3 \times S^3 \times T^4$, which is the near horizon limit of $N_5=k$ NS5-branes and $N_1$ fundamental strings. This superstring background has a six-dimensional string coupling constant fixed to $g_s = N_5/N_1$, which means that the number of fundamental strings must be large in the perturbative regime. One can argue that the spectral flow parameter must be bounded by $w\leqslant N_1$ as a consequence of the `stringy exclusion principle'~\cite{Maldacena:1998bw}. We cannot see this bound from the worldsheet description for two reasons: one is that we cannot
 properly describe multi-string states since the worldsheet theory is a first-quantized theory; the other is because 
states where $w\sim N_1$ have a strong backreaction on the background.  Said differently,  no more fundamental strings can 
`escape' to the boundary of $\textit{AdS}_3$ than the $N_1$ F1's which are part of the background geometry.  Note that by taking 
the upper bound for the spectral flow, $w=N_1$ (corresponding to the configuration, inaccessible to perturbation theory, where all the fundamental strings of the background are long strings close to the boundary), one obtains the central charge that is expected from the analysis of the D1-D5 low-energy dynamics, $c=6N_1 N_5$.


\subsection{Bosonic Warped $\textit{AdS}_3$}
\label{sec:bosonicdeformed}

We would like to repeat this calculation in the deformed WZW model (\ref{DefAs2}) in order to find which of the space-time charges survive after the deformation.\footnote{Since the  $\widehat{SL(2,\bb{R})}$ current $J^3(z)$ is a non-normalizable operator, we expect that the asymptotic charges of the background {\it will} change.} To that end,
we add a $\widehat{U(1)}$ theory, represented by a free boson $\varphi (z,\bar{z}) = \varphi(z) + \bar{\varphi} (\bar{z})$~:
\begin{equation}
J = i\partial \varphi (z)\ , \qquad \bar{J} = i \bar{\partial} \bar{\varphi}  (\bar{z}) \, ,
\end{equation}
with the OPE $\varphi (z) \varphi (0) \sim - \ln z$. This boson is compactified at radius $R$: $\varphi (z,\bar z) \cong \varphi (z,\bar z) +2  \pi R$.
The current-current deformation then takes the form
\begin{equation}
\label{currentdef}
\Delta S = -\frac{2\textsc{h}}{\pi\sqrt{k}} \int \di^2 z\, J^3 \bar{J} \, .
\end{equation}
As explained earlier, this corresponds geometrically to deforming $\textit{AdS}_3 \times S^1$ into an $S^1$ bundle over a $\textit{WAdS}_3$ base.

Appendix \ref{app:latticerotations} briefly explains the connection between these marginal deformations and rotations of the lattice that defines the torus on which the free bosons (those generating the Cartan currents) live.  The upshot is that instead of deriving new OPEs for our fields, we can use the original OPEs and construct operators with charge assignments that are determined by the rotated torus identifications (said another way, we can work in the orthonormal basis). In particular, the deformation that we consider here corresponds to an $O(2)\subset O(2,2)$ rotation that mixes the charges of $T(z)$ with those of $\bar{\varphi}(\bar{z})$.

Primary operators of the undeformed theory (\ref{freefieldprimaries}), dressed with a $U(1)$ primary with left and right momenta $p_L$ and $p_R$, lead to operators of the form
\begin{equation}
\label{defprimary}
e^{\frac{2}{\alpha_+} j\, s(z,\bar z)+i\sqrt{\frac{2}{k}}\left( m \chi (z) + \bar m \bar{\chi} (\bar z) \right) }\,
e^{-i\sqrt{\frac{2}{k}} \big( M' \hat{T} (z) + (\bar m +  \frac{kw}{2} )
\bar{\hat{T}} (\bar z)  \big) + i p_L \hat{\varphi} (z) + i p'_R
\bar{\hat{\varphi}} }\, ,
\end{equation}
where $\hat{T}$ and $\hat{\varphi}$ have canonical OPEs.
The lattice rotation gives us a relation between the new charges, the primed quantities, and charges for the original variables, the unprimed quantities.  We have the relation
\begin{subequations}
\label{chargerotation}
\begin{eqnarray}
M' &=& (m+\tfrac{kw}{2})\cos\alpha - \sqrt{\tfrac{k}{2}} \, p_R \sin\alpha \label{chargerotation_a} \, , \\
p'_R &=& p_R \cos\alpha + \sqrt{\tfrac{2}{k}} \,  (m+\tfrac{kw}{2}) \sin\alpha\, . \label{chargerotation_b}
\end{eqnarray}
\end{subequations}
The rotation angle $\alpha$ is related to the deformation parameter $\textsc{h}$ in (\ref{currentdef}) through~\cite{Israel:2003cx}
\begin{equation}
\cos^2\alpha = \frac{1}{1+2\textsc{h}^2} \, .
\end{equation}
The orthonormal basis is useful  because it allows us to have our cake and eat it, too; we get to use fields with canonical OPEs \emph{and} we can understand the new charges in terms of the unrotated charges that we already understood:
\begin{equation}
m-\mb = n \in \bb{Z} \, ,  \qquad  p_{\stackrel{L}{\text{\tiny \it R}}} = \tfrac{p}{R} \mp \tfrac{\omega R}{2} \, ,  \qquad  p,\, \omega\in\bb{Z} \, ,
\end{equation}
and, of course, $m$ and $\mb$ are further constrained depending on which representation of $\widehat{SL(2,\bb{R})}$ we are considering.

We can now immediately read off the conformal dimensions of these operators:
\begin{subequations}
\begin{eqnarray}
\Delta &=& -\frac{j(j-1)}{(k-2)} + \frac{m^2}{k} - \frac{1}{2} \Big(- \sqrt{\tfrac{2}{k}} (m+\tfrac{kw}{2}) \cos\alpha + p_R \sin\alpha \Big)^2 + \tfrac{1}{2} p_L^2 + N  \, ,   \\
\bar{\Delta} &=& -\frac{j(j-1)}{(k-2)} - w \bar{m} - \frac{kw^2}{4}
+ \frac{1}{2} \Big( p_R \cos\alpha + \sqrt{\tfrac{2}{k}}(m+\tfrac{kw}{2}) \sin\alpha \Big)^2 + \Nb \, .
\end{eqnarray}
\end{subequations}
We are interested in looking for holomorphic $(1,0)$ operators that we can use to define space-time charges. Holomorphicity imposes the
constraints  $j=0$, $\mb=0$, $w=0$, $\Nb=0$, and $p_R' = 0$.  Using (\ref{chargerotation_b}), $p_R'=0$ implies that
\begin{equation}
m = - \sqrt{\frac{k}{2}}\, p_R \cot\alpha \, .
\label{m_pr_rel}
\end{equation}
The final condition we must satisfy is reduced to
\begin{equation}
\label{bosonic-dimension}
\Delta = - \tfrac{1}{2}p_R^2 + \tfrac{1}{2} p_L^2 + N  = - \omega p + N = 1 \, .
\end{equation}
If we consider operators in the sector that has no winding around the fibered circle $\varphi$, for example, then we must set $\omega=0$ and $N=1$.  Finally, we have the condition $m-\mb=n\in\bb{Z}$; since we have set $\mb=0$, generically we can only construct a single operator with $m=p=0$.  However, if the rotation angle $\alpha$ and the compactification radius $R$ satisfy
\begin{equation}
\label{eqn:bosonic-constraint}
\sqrt{\frac{2}{k}} R \tan \alpha = \frac{2\textsc{h}R}{\sqrt{k}} \in \bb{Q} \, ,
\end{equation}
then we have an infinite set of holomorphic operators yielding consistent space-time charges.\footnote{It is also possible to construct a Virasoro algebra using a winding state of the $\widehat{U}(1)$ theory, rather than a momentum state. Then we would find a relation similar to~(\ref{eqn:bosonic-constraint}) but with the T-dual radius $\tilde{R}=\frac{2}{R}$.}  When this holds, let us write 
\begin{equation}\frac{2\textsc{h}R}{\sqrt{k}} = \frac{\mu}{\rho}\, , \end{equation} 
where $\mu$ and $\rho$ are relatively prime positive 
integers.  The space-time charges are then obtained for  $p\in\mu\bb{Z}$ and $m\in\rho\bb{Z}$.  At this point, one might wonder whether this result is immediately suspect since the physical radius of the circle could fluctuate locally; however, deformations of the radius of the circle are not normalizable and, therefore, do not correspond to finite energy excitations of the theory (similarly for deformations of $\textsc{h}$).

With these constraints in mind, we have candidate holomorphic operators, of spin $j=0$,
\begin{equation}
e^{  i \sqrt{\frac{2}{k}}\, n \left( \chi - \hat{\varphi} \tan\alpha - \hat{T} \sec\alpha \right)} \times \mathit{oscillators} \, .
\end{equation}
The form of these operators suggests defining a new orthonormal basis of free bosons:
\begin{equation}
\label{newbasis}
\tilde{T}= \hat{T} \sec\alpha + \hat{\varphi} \tan\alpha   \,  ,  \qquad  \tilde{\varphi} \equiv \hat{T} \tan\alpha + \hat{\varphi}\sec\alpha \, ,
\end{equation}
where $\tilde{T}$ is time-like and $\tilde{\varphi}$ is space-like.  Since the stress tensor and the OPEs of these free fields are the same as those used in the undeformed case, we can \emph{almost} conclude that we have a space-time Virasoro algebra constructed
out of these worldsheet holomorphic operators with modes given by
\begin{equation}
L^{\mathrm{st}}_n \stackrel{?}{=} A_0 \delta_{n,0} -  \oint \frac{\di z}{2i\pi } \bigg[ \frac{1}{2}\sqrt{\frac{k}{2}}\, i(\p \chi+ \p \tilde{T})
-  n\, \sqrt{\frac{k-2}{2}}   \p s \bigg] e^{i\sqrt{\frac{2}{k}}\, n (\chi-\tilde{T})} \, .
\end{equation}
The catch comes precisely from the constraint (\ref{eqn:bosonic-constraint}) --- when it is not satisfied, the one and only consistent operator is $L^{\mathrm{st}}_0$. Even when this constraint \emph{is} satisfied, we have the restriction $n\in\rho\bb{Z}$ which means that only some of the modes survive. This subset of Virasoro generators, by themselves, generates a full Virasoro algebra. Indeed, if we call $n \equiv \rho\hat{n}$, for $\hat{n}\in\bb{Z}$,  we note that the subset $\{ L_{\rho\hat{n}} | \hat{n}\in\bb{Z} \} \subset \{ L_n | n\in\bb{Z} \}$  generates a Virasoro algebra if we define
\begin{equation}
\label{subvirasoro}
\hat{L}_{\hat{n}} \equiv \frac{1}{\rho} L_{\rho\hat{n}} + \frac{c}{24} \big( \rho - \tfrac{1}{\rho} \big) \delta_{\hat{n},0}\, .
\end{equation}
This algebra has central charge $\hat{c} = \rho c$, where $c$ is the central charge of the algebra spanned by {\it all} the $L_n$'s.

Thus, when the condition~(\ref{eqn:bosonic-constraint}) holds
we can construct a space-time Virasoro algebra with the modes
\begin{equation}
\hat{L}^{\mathrm{st}}_{\hat n} = \frac{k\hat{p}}{4}\rho \,\delta_{\hat{n},0} - \frac{1}{\rho} \oint \frac{\di z}{2i\pi } \bigg[\frac{1}{2}\sqrt{\frac{k}{2}}\, i(\p \chi+ \p \tilde{T})   -  \rho \hat{n}\, \sqrt{\frac{k-2}{2}}   \p s \bigg] e^{i\sqrt{\frac{2}{k}}\, \rho \hat{n} (\chi-\tilde{T})} \, .
\end{equation}
Following the previous discussion, the central charge of this space-time Virasoro algebra,  in a given sector of
spectral flow $w$ and winding $\omega$ around the extra $S^1$, is given by $\hat{c}_{\mathrm{st}}=6k\hat{p}$ where
\begin{eqnarray}
\hat{p} &=& \sqrt{\frac{2}{k}} \oint \frac{\di z}{2\pi i} \big( i\p\chi - i\p\tilde{T} \big) =  \sqrt{\frac{2}{k}}\oint \frac{\di z}{2i\pi } (i\p \chi -i \sec \alpha\,\p \hat{T} -i \tan \alpha \, \p\hat{\varphi} )\nonumber\\
&=& w + \sqrt{\frac{2}{k}}\tan \alpha (p_R - p_L) = w +\frac{\mu}{\rho} \omega
\end{eqnarray}
and where we used the relation~(\ref{eqn:bosonic-constraint}) to simplify. Therefore, when there exists a second Virasoro algebra,  we obtain $\hat{c}_{\mathrm{st}}= 6k (\rho w + \mu \omega )$.  

To interpret this result, notice that in the time-like warped {\it AdS}$_3$ background, we still have long string solutions in sectors of non-zero spectral flow \cite{Israel:2003cx}. As in {\it AdS}$_3$, the relation between space-time and worldsheet time is $t = w \tau$,\footnote{Although the manifold now 
has CTCs, we still call time the coordinate corresponding to the isometry $J^3_0 + \bar J^3_0$ of the worldsheet theory.} independent of the winding number $\omega$ around the circle. Thus, as in the {\it AdS}$_3$ computation, the orientation of the integration contour that gives the central extension should be reversed for negative spectral flow parameter $w$. We expect that the 
left space-time central charge is, then, given by 
\begin{equation}\left\{ 
\begin{array}{lcc} \hat{c}_{\mathrm{st}}= 6k (\rho w + \mu \omega )\, , &\quad w>0 \, , \\
 \hat{c}_{\mathrm{st}}= -6k (\rho w + \mu \omega )\, , &\quad  w<0\, .
\end{array} \right.
\label{wcentralcharge}
\end{equation}
Of course, on the antiholomorphic side nothing has been deformed so we always have a commuting space-time Virasoro algebra with $\bar{c}_{\mathrm{st}} = 6k|w|$.  Again, these results apply in a sector of given spectral flow $w$ and winding $\omega$, see the discussion below (\ref{centralcharge}).  We will discuss the implications of~(\ref{wcentralcharge}) in the next subsection.

So far, we have ignored another family of potential space-time charges at our disposal which are made from worldsheet primaries of the form $i\p \tilde{\varphi}\, e^{i\sqrt{\frac{2}{k}} n (\chi-\tilde{T})}$. Let us consider the case where the affine $U(1)$ used in the 
construction is at level $k_g$.  
In this case, we obtain a space-time $\widehat{U(1)}$ algebra when (\ref{eqn:bosonic-constraint}) holds, with generators
\begin{equation}
\hat{J}^{\mathrm{st}}_{\hat{n}} =  \oint \frac{\di z}{2i\pi } \,  \sqrt{\frac{k_g}{2}} i\p \tilde{\varphi}\, e^{i\sqrt{\frac{2}{k}} \rho\hat{n} (\chi-\tilde{T})}\, ,
\end{equation}
and only $\hat{J}^{\mathrm{st}}_0$ when it fails.  The level $k^{\mathrm{st}}_g$ of this space-time affine $U(1)$ is
\begin{equation}
\frac{k^{\mathrm{st}}_g}{2} = [ \hat{J}^{\mathrm{st}}_1 ,\,  \hat{J}^{\mathrm{st}}_{-1} ] =   \rho \frac{k_g}{2} \oint \frac{\di z}{2i \pi}  \sqrt{\frac{2}{k}}(i\partial \chi - i\partial \tilde{T}) \, ,
\end{equation}
so we obtain
\begin{equation}
\left\{
\begin{array}{lccc}
k^{\mathrm{st}}_g &=& k_g (\rho w + \mu \omega )\, , &\quad  w>0 \, , \\
k^{\mathrm{st}}_g &=& -k_g (\rho w + \mu \omega ) \, , & \quad  w<0 \, .
\end{array} \right.
\end{equation}
In a similar way, {\it any} current algebra at level $k_{i}$ from the internal CFT, $\mathcal{M}$, will be promoted to an affine symmetry in space-time at level $\pm k_i (\rho w + \mu \omega )$.

To summarize, we find a somewhat peculiar result when \wads~is constructed in string theory as the base of an $S^1$ fibration: we can tune the warping parameter $\textsc{h}$ and the $S^1$ radius $R$ so that when $\frac{2\textsc{h}R}{\sqrt{k}}$ is rational, we have two space-time Virasoro algebras (generically with different central charges), whereas when it is irrational we have only one Virasoro algebra along with $L_0^{\mathrm{st}}$ and an additional global $U(1)$ symmetry, $J_0^{\mathrm{st}}$.  If the reader worries this is somehow an artifact of the free field calculation, we have reproduced the above calculation in appendix \ref{app:bosonicbrst} without reference to free fields. Furthermore, in section \ref{sec:ASG} we perform the asymptotic symmetry analysis on the four-dimensional geometry and find precisely the same result from supergravity, though only in the sector without winding around the $S^1$, $\omega=0$.


\section{Space-time Symmetry Algebra II:~~Heterotic and Type II Strings}\label{superstring}

Next, we would like to find what space-time symmetries for heterotic or type IIA/B superstrings on $\textit{WAdS}_3 \times S^1 \times T^3 \times S^3$ can be realized in the worldsheet CFT. In particular, one might wonder whether the symmetry enhancement found in the bosonic case gives rise to a full extra superconformal algebra in space-time.  To answer this, we'll consider a current-current deformation involving the elliptic/time-like holomorphic current of the left $\mathcal{N}=1$ $\widehat{SL(2,\mathbb{R})}$ algebra (hence chosen as the supersymmetric side in heterotic) with the antiholomorphic current $i \bar{\partial} \varphi$ from the $S^1$. Thus, this analysis is common to type II and heterotic models, with another possibility briefly considered in section \ref{sec:het-bundle}. 

Consider, then, a $(1,0)$ or $(1,1)$ $SL(2,\mathbb{R})\times U(1)$  WZW model. The holomorphic $\widehat{SL(2,\mathbb{R})}$ current algebra at level $k$ is generated by
\begin{equation}
\mathcal{J}^A (z) \equiv J^A (z) - \frac{i}{2} \epsilon^A_{\ph{A}BC} : \psi^B \psi^C : (z) \, ,
\end{equation}
where the $J^A$ are generators of a bosonic $\SL_{k-2}$ and the $\psi^A$ are three free Majorana-Weyl fermions, normalized as $\psi^A \psi^B \sim \frac{1}{z}\eta^{AB}$, with metric $\eta_{33} = -1$, $\eta_{+-}=\frac{1}{2}$, $\epsilon^3_{\ph{3}12}=1$.
This theory admits an $\mathcal{N}=(1,0)$ supersymmetry on the worldsheet with supercurrent given by
\begin{equation}
G =  \sqrt{\frac{2}{k}} \Big( -\psi^3 \mathcal{J}^3 + \frac{1}{2} \psi^+ J^- + \frac{1}{2} \psi^- J^+ \Big) \, .
\end{equation}
In $(1,0)$ superspace notation, the current multiplet is $\Xi^A = \sqrt{\frac{k}{2}} \,\psi^A + \theta \mathcal{J}^A$. Thus, we can deform the space-time action as before while preserving worldsheet supersymmetry by adding an operator of the form
\begin{equation}
\label{susydef}
\Delta S = -\frac{2\textsc{h} }{\pi\sqrt{k}} \int \di^2 z \, \di \theta\, \, \Xi^3 \bp\varphi = -\frac{2\textsc{h}}{\pi\sqrt{k}} \int \di^2 z  \, \mathcal{J}^3 \bp\varphi \, .
\end{equation}

As the underlying lattice rotation acts on the lattice of $\mathcal{J}^3$ and \emph{not} on the purely bosonic current $J^3$, it is useful to split the $(1,0)$ WZW model in terms of non-compact parafermions which correspond to the \emph{superconformal coset} CFT $SL(2,\mathbb{R})/U(1)$. Thus, we write
\begin{equation}
\mathcal{J}^3 = i\sqrt{\frac{k}{2}} \p \mathfrak{T} \, ,
\end{equation}
where $\mathfrak{T}$ is a time-like boson.  The SCFT $SL(2,\mathbb{R})/U(1)$ actually admits an $\mathcal{N}=(2,0)$ superconformal algebra; defining $Q\equiv \sqrt{\frac{2}{k}}$, we can use free fields to write the generators as\footnote{This description of the SCFT is usually used in the literature to describe the (mirror) $\mathcal{N}=2$ Liouville theory, see {\it e.g.}~\cite{Hosomichi:2004ph}.}
\begin{subequations}
\label{n2algebra}
\begin{eqnarray}
T&=& - \tfrac{1}{2} \p s \p s - \tfrac{1}{2} \p X \p X - \tfrac{1}{2} \p H_1 \p H_1 + \frac{Q}{2} \p^2 s  \, ,  \\
 G^{\pm} &=&   (i\partial X + Q i \p H_1 \pm \partial s  ) e^{\mp i H_1} \label{n2supercur}  \, , \\
J_R &=& -i \p H_1 +  Q i \p X  \, ,
\end{eqnarray}
\end{subequations}
with the bosonization
\begin{equation}
\Psi^{\pm} \equiv \frac{\Psi^X \pm i \Psi^s}{\sqrt{2}} \equiv e^{\mp i H_1}\, .
\label{bos1}
\end{equation}
We also bosonize
\begin{equation}
\frac{1}{\sqrt{2}} (\psi^\varphi \pm \psi^3) = e^{\pm H_2}\, ,
\label{bos2}
\end{equation}
with $H_2$ time-like.

We can use a lattice rotation to understand how a primary in the deformed theory relates to one in the undeformed theory. We consider primaries in the left NS sector, of the form
\begin{equation}
\label{vertexsusy}
\V_{j m \bar m p_L p_R}^w = e^{Q j s + iQ(m X + \bar m \bar X - (m-\frac{k}{2}w) \mathfrak{T}
- (\bar m-\frac{k}{2}w) \bar{\mathfrak{T}})} e^{i ( p_L \varphi + p_R \bar \varphi)} e^{ \frac{1}{2}(i\sigma_1 H_1 + \sigma_2 H_2)}
\times V_{\text{other}} (z,\bar z)
\end{equation}
with $\sigma_1$ and $\sigma_2$ even --- the Ramond sector of the theory, where $\sigma_1$ and $\sigma_2$ are odd, will be considered below.
The corresponding deformed operator comes from an $O(2)$ rotation between
the $\mathfrak{T}$ and $\bar{\varphi}$ lattices (respectively left and right). We achieve the deformed operator by the replacements
\begin{subequations}
\begin{eqnarray}
m-\tfrac{kw}{2} &\rightarrow& (m-\tfrac{kw}{2})\cos\alpha - Q^{-1} p_R \sin\alpha \, ,  \\
p_R &\rightarrow& p_R \cos\alpha + Q \big(m-\tfrac{kw}{2}\big)  \sin\alpha \, .
\end{eqnarray}
\end{subequations}
As before, the hatted fields $\hat \varphi$ and $\hat{\mathfrak{T}}$ correspond to the orthonormal basis and, thus, maintain canonical OPEs along the deformation line.

The conformal dimensions of spinless operators ($j=0$) with no spectral flow ($w=0$) in the deformed theory will be
\begin{subequations}
\begin{eqnarray}
\Delta &=& \frac{m^2}{k} - \frac{1}{2} \Big(  Q m\cos\alpha -  p_R \sin\alpha \Big)^2 + \frac{\sigma_1^2+\sigma_2^2}{8} +
\frac{p_L^2}{2} \, ,   \\
\bar{\Delta} &=& \frac{1}{2} \Big(  p_R \cos\alpha + Q m \sin\alpha \Big)^2 \, .
\end{eqnarray}
\end{subequations}
Again, our goal will be to find integrated 
holomorphic $(1,0)$ operators that we can use to build the space-time Virasoro charges. In the usual way, 
we will  look for holomorphic $(\frac{1}{2},0)$ operators ${\cal O}_n$ that allow us to write $L_n = \{G_{-\frac{1}{2}},{\cal O}_n\}$. Then the $L_n$'s  will be BRST-invariant if $\{G_{\frac{1}{2}},\mathcal{O}_n \} = 0$.

Setting $\bar{\Delta} = 0$ implies that the right-moving theory is in the NS sector (for both type II and heterotic). To be holomorphic, the operators must satisfy
\begin{equation}
\label{eqn:het-constraint}
m = - \frac{\cot\alpha}{Q} p_R \, .
\end{equation}
Since $m-\bar m = n \in \mathbb{Z}$ in the NS-NS sector and since $\bar m =0$, we arrive at the same constraint (\ref{eqn:bosonic-constraint}) as in the bosonic case.  When this condition is satisfied, then conformal dimension becomes
\begin{equation}
\label{margineqn}
\Delta = \frac{\sigma_1^2+\sigma_2^2}{8} + \frac{1}{2}\big( p_L^2 - p_R^2 \big) =  \frac{\sigma_1^2+\sigma_2^2}{8} - \omega p \, .
\end{equation}
As before, we consider winding-less states ($\omega=0$).  Then the possible building blocks for the
space-time supercharges satisfy
\begin{equation}
\sigma_1^2 +\sigma_2^2 = 4\, ,
\end{equation}
which is solved with $(\sigma_1, \sigma_2) = (\pm 2, 0)$ or $( 0,\pm 2)$.

As in the  bosonic case, we expect that a special role is played by the rotated fields $\tilde{\mathfrak{T}}$ and $\mathfrak{F}$, defined through a change of orthonormal basis by
\begin{equation}
\label{newbasisbos}
\tilde{\mathfrak{T}}= \hat{\mathfrak{T}} \sec\alpha + \hat{\varphi} \tan\alpha   \,  ,  \qquad  \mathfrak{F} \equiv \hat{\mathfrak{T}} \tan\alpha + \hat{\varphi}\sec\alpha \, .
\end{equation}
Naturally, one can define a similar change of basis for the fermionic superpartners so that the supercurrent is preserved:
\begin{equation}
\label{newbasisfer}
\Psi^3= \psi^3 \sec\alpha + \psi^\varphi \tan\alpha   \, ,  \qquad  \Psi^{\mathfrak{F}} 
\equiv  \psi^\varphi \sec\alpha + \psi^3 \tan\alpha  \, .
\end{equation}
Thus, the supercurrent of the theory can be written as
\begin{equation}
\label{deforsupercur}
\hat{G} =\frac{1}{\sqrt{2}}( G^+ + G^- ) - i \Psi^3 \partial \tilde{\mathfrak{T}} + i \Psi^{\mathfrak{F}} \partial \mathfrak{F} \, ,
\end{equation}
with $G^{\pm}$ as in (\ref{n2supercur}).

We don't expect the fermionic field $\Psi^{\mathfrak{F}}$ to be part of the construction of the space-time Virasoro algebra since it's orthogonal to the superpartner of $\tilde{\mathfrak{T}}$. Therefore, the candidate Virasoro generators are of the form
\begin{equation}
L^{\mathrm{st}}_n = \oint \frac{\di z}{2i\pi}  \, \Big\{ G_{-\frac{1}{2}}\, , \, ( b^-_n \Psi^- + b^+_n \Psi^+ +  b^3_n \Psi^3 ) e^{i Q n (X-\tilde{\mathfrak{T}})} \Big\} \, .
\end{equation}
As before, because of the constraint (\ref{eqn:bosonic-constraint}) these operators exist only for $n \equiv 0 \mod \rho$ when (\ref{eqn:bosonic-constraint}) is satisfied, and only for $n=0$ otherwise.
Demanding that $\mathcal{O}$ is a superconformal primary leads to the constraint
\begin{equation}
(n+1)b^-_n + (n-1)b^+_n + \sqrt{2} n b^3_n = 0 \, .
\end{equation}
Moreover, for $n \neq 0$ this operator is defined modulo exact terms which take the form
\begin{equation}
\{ G_{-\frac{1}{2}},  e^{i Q n (X-\tilde{\mathfrak{T}})}  \} = Q n \left(\frac{\Psi^+ + \Psi^-}{\sqrt{2}} - \Psi^3 \right)  e^{i Q n (X-\tilde{\mathfrak{T}})} \, ,
\end{equation}
corresponding to the equivalence
\begin{equation}
(b^-_n,b^+_n,b^3_n) \cong (b^-_n + \lambda,b^-_n +\lambda,b^3_n  -\sqrt{2} \lambda) \, , \quad \forall ~\lambda \in \mathbb{R} \, .
\end{equation}
Using this gauge freedom to set $b^3_n = 0$ for $n\neq0$, $\mathcal{O}$ will be a superconformal primary if we choose $b_n^\pm = -\frac{1}{\sqrt{2}Q}(1\pm n)$ for $n\neq 0$.  This yields space-time Virasoro generators
\begin{equation}
\label{virasusy}
L^{\mathrm{st}}_{\hat{n}} = \frac{k\hat{p}}{4} \big( \rho + \tfrac{1}{\rho} \big)\delta_{\hat{n},0} - \frac{1}{Q \rho} \oint \frac{\di z}{2i\pi}
\left[ i\partial X - \rho  \hat n  \partial s  +  Q \rho \hat n \Psi^X  \Psi^3  - i Q \rho^2 \hat n^2 \Psi^s( \Psi^X- \Psi^3 )
\right] e^{i Q \rho \hat{n} (X-\tilde{\mathfrak{T}})} \, ,
\end{equation}
where
\be
\hat{p} \equiv iQ\oint\frac{\di z}{2i\pi} \big( \p X - \p\tilde{\mathfrak{T}}\big) = w + \frac{\mu}{\rho}\omega \, .
\ee
As in the bosonic case, the central charge of this algebra is
\begin{equation}\left\{ 
\begin{array}{lcc} \hat{c}_{\mathrm{st}}= 6k (\rho w + \mu \omega ) \, , &\quad w>0 \, , \\
 \hat{c}_{\mathrm{st}}= -6k (\rho w + \mu \omega ) \, , &\quad  w<0 \, .
\end{array} \right.
\label{wcentralchargesusy}
\end{equation}
Again, we can also construct an affine $U(1)$ algebra in space-time from the extra free boson $\mathfrak{F}$ using operators of the form $\{ G_{-\frac{1}{2}}\, , \, \Psi^{\mathfrak{F}} e^{iQ \rho \hat n (X- \tilde{\mathfrak{T}})} \}$, as well as other affine algebras in space-time coming from the internal CFT (for instance, $SU(2)_k \times U(1)^3$).

We would like to interpret this result in parallel to the discussion around~(\ref{centralcharge}). The time-like 
warped {\it AdS}$_3$ space-time, for any value of the warping parameter $\textsc{h}$, is known to have CTCs 
passing through each point.  The worldsheets of long fundamental strings  have their radius increasing linearly with time and eventually must wrap these CTCs. The analysis of space-time symmetries that we gave above is relevant for the worldvolume 
conformal field theory on these long strings, hence we expect to find some holographic signatures of a pathological behavior. From equation~(\ref{wcentralchargesusy}) above, one can notice immediately two important facts:
\begin{enumerate}
\item $c_{\mathrm{st}}-\bar{c}_{\mathrm{st}} \, \, \slash \!\!\!\!\!\! \equiv 0 \mod 24$, generically, implying that the purported dual CFT is not modular invariant and, hence, inconsistent.
\item Nothing seems to prevent $c_{\mathrm{st}}$ from becoming arbitrarily negative for large negative winding numbers $\omega$, spoiling unitarity of this potential CFT. This would correspond to the worldvolume theory on $w$ long strings 
that wind around the circle.\footnote{As  the string coupling goes like $g_s \sim 1/N_1$, regardless of the winding number around the $S^1$, we expect that the `stringy exclusion principle' still gives the bound $w \leq N_1$ here for any value of $\omega$.} 
\end{enumerate}

These aspects deserve further study and will hopefully be clarified if a brane construction of this background is found. It's also worth mentioning that space-like $\textit{WAdS}_3$ does not suffer from the same pathologies as its time-like cousin, so we expect better behavior of the space-time central charge in that case.  Unfortunately, the analysis in the hyperbolic/space-like basis of $\SL$ is significantly more difficult and, thus, left for future work.

\subsection{Space-time Supersymmetry}
\label{sec:space-time-susy}

We have seen that for rational values of $\frac{2\textsc{h}R}{\sqrt{k}}$ it is possible to construct a space-time Virasoro algebra from the left-moving worldsheet CFT, though it has a different central charge than the space-time Virasoro algebra arising from the right-moving sector.  The next natural question to ask is whether this space-time Virasoro algebra admits a supersymmetric extension at rational values of $\frac{2\textsc{h}R}{\sqrt{k}}$ (of course, in the type II case we always have at least one space-time super-Virasoro algebra from the right-moving sector).

In order to answer this question, we search for space-time supercharges in the R-NS sector of the worldsheet CFT. Let us first remind ourselves of the form of the space-time supercharges in the undeformed background, $\textit{AdS}_3\times U(1) \times \mathcal{M}/U(1)$, where the CFT $\mathcal{M}/U(1)$ has an $\mathcal{N}=2$ superconformal symmetry and central charge $c_m = 9-\frac{6}{k}$.
 We use the notation
\begin{equation}
J = i \partial U \, , \qquad J_R^{\mathcal{M}/U(1)} = \sqrt{3-\tfrac{2}{k}}\, i \p V \, .
\end{equation}
In this case, the operators
\begin{equation}
G^{\pm,\mathrm{st}}_r = \oint \frac{\di z}{2i\pi}\, e^{-\phi_L/2}  e^{r \big(-iH_1 + iQ(X-\mathfrak{T})\pm H_2 \big) \pm \frac{i}{2} \big( \sqrt{3-\frac{2}{k}}\, V-QU \big) } \, , \qquad r = \pm \frac12 \, ,
\end{equation}
(where the bosonizations were given in (\ref{bos1}) and (\ref{bos2}) and $\phi_L$ is a superghost) are BRST invariant and mutually local and, therefore, act as space-time supercharges on the physical string Hilbert space.  The other modes of $G^{\pm,\mathrm{st}}$, for which $r\neq \pm\frac{1}{2}$, are then obtained by acting with $L^{\mathrm{st}}_n$.

Now, let's search for potential space-time supercharges in the deformed case.  As a first attempt, let's identify the distinguished $U(1)$ symmetry of the undeformed case with the $U(1)$ used in the deformation, {\it i.e.} let $U=\varphi$.
Next, bosonize the rotated fermions of the deformed model (\ref{newbasisfer}),
\begin{equation}
e^{\pm \mathfrak{H}} \equiv \frac{1}{\sqrt{2}} (\Psi^{\mathfrak{F}} \pm \Psi^3) \, .
\end{equation}
Then it is natural to guess that the space-time supercharges take the form
\begin{equation}
G^{\pm,\mathrm{st}}_r \stackrel{?}{=} \oint \frac{\di z}{2i\pi}\, e^{-\phi_L/2}  e^{r \big( -iH_1+ iQ(X-\tilde{\mathfrak{T}})\pm \mathfrak{H} \big) \pm \frac{i}{2} \big( \sqrt{3-\frac{2}{k}}\, V-Q\mathfrak{F} \big) } 
\label{superchargescand}
\end{equation}
since the OPEs and the supercurrent (\ref{deforsupercur}), expressed in terms of the rotated variables in the deformed case, are the same as those in the undeformed case expressed in terms of the unrotated variables, thus guaranteeing BRST invariance.

So we have identified an operator with the OPEs that a space-time supercurrent should have, but we have to check that it's actually in the spectrum of the deformed model.  From the definition of the rotated boson $\mathfrak{F}$, (\ref{newbasisbos}), and from the definition of the vertex operators in (\ref{vertexsusy}), we arrive at the identifications
\begin{subequations}
\begin{eqnarray}
m\cos \alpha - Q^{-1}p_R \sin \alpha  &=& r \sec \alpha \pm \frac12 \tan \alpha \, ,\\
p_L &=&  -Q r\tan \alpha \mp \frac12 Q \sec \alpha \, , \\
p_R \cos \alpha + Qm \sin \alpha &=&0 \, .
\end{eqnarray}
\end{subequations}
Eliminating $p_R$ between the first and third equations leads to
\begin{equation}
m = r \mp \frac{1}{2} \sin \alpha \, ,
\end{equation}
but since we are in the R-NS sector, where $\bar{m}=0$ implies that $m\in\mathbb{Z} + \tfrac{1}{2}$, and since conformality requires $r=\pm\frac{1}{2}$, this operator can only be part of the spectrum if $\alpha \equiv 0 \mod \pi$, {\it i.e.} in $\textit{AdS}_3$.  

Even in the undeformed case, this construction is not possible for a generic $U(1)$ lattice (at least, without any shift orbifold mixing it with another CFT) since one would need $p_L = \mp Q/2$ and $p_R = 0$.  In the case of $\textit{AdS}_3 \times S^3 \times T^4$, this is accomplished by choosing the $U(1)$ to be the Hopf fiber of the $S^3$.  This suggests another construction by using {\it another} $U(1)$ in order to construct the space-time supercharges.  To be concrete, consider deforming the model
\begin{equation}
SL(2,\mathbb{R})_k \times U(1)_\varphi \times \frac{\left[ \frac{SU(2)_k}{U(1)_k}   \times U(1)_k \right]}{\mathbb{Z}} \times U(1)^3 \, ,
\end{equation}
using $\varphi$ to generate the warping, as before, while building the space-time supercharges with the $U(1)_k$ of the 
$SU(2)_k$ algebra, as is done in the undeformed geometry $\textit{AdS}_3\times S^3 \times T^4$.  Let us denote the $U(1)_k$ current by $i\sqrt{\frac{k}{2}} \,\partial Z$ and bosonize
\begin{equation}
e^{\pm H_3} = \frac{1}{\sqrt{2}} (\psi^Z \pm \Psi^3) \, .
\end{equation}
As above, the idea is to use the {\it rotated} bosons and fermions $(\tilde{\mathfrak{T}},\Psi^3)$ and $(\mathfrak{F},\Psi^{\mathfrak{F}})$ in order to work with canonical OPEs.  This would suggest defining supercharges as
\begin{equation}
\label{superchargecandidate}
G^{\pm,\mathrm{st}}_r = \oint \frac{\di z}{2i\pi}\, e^{-\phi_L/2}  e^{r \big(-iH_1+ iQ(X-\tilde{\mathfrak{T}})\pm H_3\big) \pm \frac{i}{2} \left( \sqrt{3-\frac{2}{k}}\, V- Q Z \right) } \, , \qquad r = \pm \frac12 \, .
\end{equation}
These operators have $m=r$ and
\begin{equation}
p_L = p_R = - Qr \tan \alpha = -\frac{2\textsc{h}}{\sqrt{k}}=- \frac{\mu}{\rho R}\, ,
\end{equation}
where we have used (\ref{eqn:bosonic-constraint}).  

Therefore, these physical space-time supercharges can be constructed in the worldsheet theory when $\rho=1$, meaning that in these cases we actually have a space-time super-Virasoro algebra (two, in the case of type II strings).  On the other hand, for generic values of $\frac{2\textsc{h}R}{\sqrt{k}}$ there is no space-time supersymmetry generated by the left-moving side of the worldsheet.  In type II constructions, this implies that half of space-time supersymmetry is broken, while for heterotic constructions this implies that all supersymmetry is broken (though in the heterotic case, we can instead deform the right-movers and preserve all of space-time supersymmetry).  We will find the same results from supergravity in section \ref{sec:ASG}.

\subsection{Heterotic Strings With a Gauge Bundle}
\label{sec:het-bundle}
There is one more case to consider: in heterotic string theory, instead of realizing the $\textit{WAdS}_3$ background 
with an $S^1$ fibration we can use a ten-dimensional Abelian gauge field; in other words, we take $\bar{\varphi}$ to be a right-moving chiral boson. 

In order to understand the space-time symmetries of the solution, we redo the same analysis as before except that we have to impose the constraint $p_L=0$.  Recall that charges corresponding to the space-time Virasoro algebra should be built from operators satisfying
\begin{equation}
\Delta = \frac{\sigma_1^2+\sigma_2^2}{8} - \frac{1}{2} p_R^2 + N =  \frac12 \, ,
\end{equation}
where $p_R$ is in the $(E_8\times E_8)_1$ or $Spin(32)_1 /\mathbb{Z}_2$ lattice, see~(\ref{margineqn}). 
Note that compared to the previous analysis, $H_2$ refers to a different bosonization of $\psi^3$ than before~(\ref{bos2}) since $\psi^\varphi$ is not part of the heterotic model. 

Considering that worldsheet fermions of the `internal' CFT do not appear in the vertex operators we are looking for, 
the left GSO projection imposes that $\sigma_1+\sigma_2  \equiv 2 \mod 4$. Comparing 
with the discussion below~(\ref{margineqn}), it is clear that one cannot use the same construction as 
for the case with the $S^1$ fibration. Indeed, to get a generator $L^{\mathrm{st}}_n$ one would need to solve the constraint
\begin{equation}
n^2 \tan^2 \alpha =  k N \, .
\end{equation}
If the deformation parameter is such that $\tan^2 \alpha$ is rational, one cannot exclude {\it a priori} that 
such operators --- which involve some oscillator modes --- can be physical for some values of $n$. However, due to the proliferation of oscillator 
combinations for a generic level $N$, it seems impossible to give a general proof that there is no 
new space-time algebra, although it seems highly probable to us that the $U(1)$ is not extended to a second Virasoro.  The best that we can say is that similar operators exist in the case the $U(1)$ arises from a geometric $S^1$, see (\ref{bosonic-dimension}): in that case, we have a better handle on the space-time analysis of the asymptotic symmetries (see section \ref{sec:ASG}) and find no evidence that a space-time Virasoro algebra can be built from operators with $N>1$.  We expect that the result is the same in this case.


\section{Space-time Symmetry Algebra III:~~Supergravity Analysis}
\label{sec:ASG}


In previous section, we analyzed the space-time symmetries of certain $\textit{WAdS}_3$ backgrounds using worldsheet techniques.  In this section, we'll explore the same question from the supergravity point of view.  First, we look at the asymptotic bosonic symmetries of these backgrounds, then we move on to consider supersymmetries.

\subsection{Asymptotic Bosonic Symmetries}

To analyze the asymptotic symmetries, let us return to our ten-dimensional background, \re{10dBG}:
\begin{equation} \label{10dBG-2}
\di s^2 = \di s^2_4 + \di s^2_{S^3} + \di s^2_{T^3} \, ,
\end{equation}
where
\begin{equation}
\di s^2_4 = k (-\cosh^2 \!\varrho \di t^2 + \di \!\varrho^2 + \sinh^2 \!\varrho \di\phi^2) - 2 \textsc{h} \, \sqrt{k} \, \di\varphi (\cosh^2\!\varrho \di t + \sinh^2\!\varrho \di \phi) + \tfrac{1}{2} \di\varphi^2 \, ,
\end{equation}
$\varrho,~t,$ and $\phi$, are the usual global coordinates on $\textit{AdS}_3$, and $\varphi \in [0,2\pi R)$.\footnote{As in section \ref{WAdS3}, we work in units where $\alpha'=1$.}  The 3-sphere has radius $L_S$ and $T^3$ has volume denoted by $V_{T^3}$. The equations of motion impose that the $\textit{AdS}_3$ and $S^3$ have radii $\sqrt{k}$. 

Solving the Killing equations corresponding to the  four-dimensional part of the metric yields the following vector fields:
\begin{subequations}
\begin{eqnarray}
& &\left\{ \begin{array}{ccl}
\bar l_0 &=& \frac{1}{2} (\p_t +\p_\phi) \, , \\
\bar l_1 &=& \frac{1}{2} e^{i (t+\phi)}  (\tanh\!\varrho \p_t - i \p_\varrho+ \coth \!\varrho \p_\phi) \, , \\
\bar l_{-1} &=& \frac{1}{2} e^{-i (t+\phi)}  (\tanh\!\varrho \p_t + i \p_\varrho + \coth\!\varrho \p_\phi) \, , 
\end{array} \right. \\
& &\left\{ \begin{array}{ccl} l_0 &=& \frac{1}{2} (\p_t - \p_\phi) \, , \\
 l_1 &=& \frac{1}{2} e^{i \big(t-\phi + \frac{2 \textsc{h}}{\sqrt{k}} \varphi \big)}  (\tanh\!\varrho \p_t - i \p_\varrho - \coth\!\varrho \p_\phi) \, , \\
 l_{-1} &=& \frac{1}{2} e^{-i \big(t-\phi + \frac{2 \textsc{h}}{\sqrt{k}} \varphi\big)}  (\tanh\!\varrho \p_t + i \p_\varrho - \coth\!\varrho \p_\phi) \, ,  \end{array} \right. \\
&& \quad \ t_0 \quad \ = \ \p_\varphi \, ,
\end{eqnarray}
\end{subequations}
forming an $\ls \oplus \ls \oplus u(1)$ algebra. However, since $\varphi$ has periodicity $2\pi R$,  $ l_1$ and  $l_{-1} $ are globally-defined vector fields only if
\begin{equation} \label{condKill}
\frac{2 \textsc{h} R}{\sqrt{k}} \in \ZZ \, ,
\end{equation}
which coincides precisely with~\re{eqn:bosonic-constraint} for $\rho=1$.

By studying the asymptotic symmetries of the above ten-dimensional background, one naturally expects the $\ls \oplus \ls$ part of the exact symmetries to be extended into the sum of two Virasoro algebras, as is familiar from $\textit{AdS}_3$ gravity~\cite{Brown:1986nw}. We will refrain from performing a full analysis --- this would involve determining a set of consistent boundary conditions that include both the background \re{10dBG-2} as well as finite temperature excitations of it obtained by discrete identifications, {\it e.g.} \cite{Banados:2005da, Compere:2009zj} --- and instead will make a natural guess for the form of the Virasoro algebra generators, leaving the full analysis for future work.  To that end, consider the following large diffeomorphisms:
\begin{eqnarray}
\bar l_n &=& \frac{1}{2} e^{i n (t+\phi)} (\tanh\!\varrho \p_t - i n \p_\varrho + \coth\!\varrho \p_\phi) \, , \\
l_n &=& \frac{1}{2} e^{i n \big(t-\phi + \frac{2 \textsc{h}}{\sqrt{k}}\varphi\big)} (\tanh\!\varrho \p_t - i n \p_\varrho - \coth\!\varrho \p_\phi) \, .
\end{eqnarray}
These satisfy the Virasoro algebra, without central term,
\begin{equation}
 i [l_m,l_n] = (m-n) l_{m+n} \, , \qquad  i [\bar l_m,\bar l_n] = (m-n) \bar l_{m+n} \, .  \nonumber
\end{equation}

Assuming that these diffeomorphisms yield finite, asymptotically conserved, integrable charges defined on an appropriate phase space that includes the background~\re{10dBG-2}, and assuming that they are represented through the Poisson bracket~\re{poissonbracket}, we can compute the central term of the algebra using \re{eq:cc}.\footnote{The Mathematica code implementing the expressions for the charges displayed in the appendix can be downloaded from the homepage of G. Comp\`ere: http://www.physics.ucsb.edu/$\sim$gcompere/}  Doing so, we find central terms
\begin{equation}
\label{centralterms}
 K_{l_m,l_n} = \frac{c}{12} m (m^2 - \alpha_0) \delta_{m+n,0} \, ,  \qquad  K_{\bar l_m, \bar l_n} = 
\frac{\bar c}{12} m (m^2 - \alpha_0) \delta_{m+n,0} \, ,
\end{equation}
where
\begin{equation} \label{c10}
c = \bar c = \frac{6 \pi^3 \sqrt{1 + 2\textsc{h}^2} \sqrt{\frac{k}{2}} R L_S^3 V_{T^3}}{G_{10}}
\end{equation}
and $G_{d}$ is the $d$-dimensional Newton's constant. When reduced to 4 or 3 dimensions, with $G_{10} = 2 \pi^2 L_S^3 V_{T^3} G_4 = 2\sqrt{2} \pi^3 R L_S^3 V_{T^3}   G_3$, we find
\begin{equation}
   c = \bar c = \frac{3  \sqrt{\frac{k}{2}} \sqrt{1 + 2\textsc{h}^2}  \pi R }{G_4} = \frac{3  \sqrt{k} \sqrt{1 + 2\textsc{h}^2}}{2 G_3}.
\end{equation}
A few comments are in order.  In~\re{centralterms}, $\alpha_0$ is a constant which can be set to any value by a redefinition of the $l_0$ and $\bar l_0$ generators, {\it i.e.} by fixing the energy of the background, so we can make the canonical choice of $\alpha_0=1$. When the squashing parameter goes to zero, the three-dimensional central charges coincide with the Brown-Henneaux charges as expected (the same holds for space-like warped $\textit{AdS}_3$, see \cite{Compere:2008cw}). However, the central charges depend not only on the parameters present in the Lagrangian ({\it e.g.} $k$ and $G_3$) but also on a parameter of the solution, $\textsc{h}$. This is reminiscent of the Kerr/CFT correspondence where the central charge depends on the angular momentum $J$ of the solution, but it is in contrast to the situation in TMG where the squashing parameter of the solution is entirely fixed in terms of the gravitational Chern-Simons coupling. 

It's also important to notice that the condition \re{condKill} can be relaxed to
\beq \label{condKill2}
 \frac{2 \textsc{h} R}{\sqrt{k}} \equiv \frac{\mu}{\rho} \in \mathbb Q \, ,
\eeq
where $\mu,\rho\in\ZZ$ are relatively prime, if we only keep the subset of $l_n$ generators where $n\in\rho\ZZ$. Indeed, the vector fields $ l_n^\prime \sim \frac{1}{\rho}  l_{\rho n}$ (see \re{subvirasoro}) are globally defined and will generate a Virasoro algebra with central charge $c' = \rho c$ \cite{Banados:1998wy,Martinec:2001cf,Martinec:2002xq}.

Previous treatments of the asymptotic symmetries of $\textit{WAdS}_3$ only revealed the existence of a single Virasoro algebra, supplemented with current algebras \cite{Compere:2007in, Compere:2008cv, Compere:2009zj},\footnote{Although, a second Virasoro algebra can be constructed out of a residual current algebra by means of a Sugawara-like construction, {\it e.g.} \cite{Blagojevic:2009ek}} which has some tension with the proposal that space-like stretched $\textit{AdS}_3$ solutions of TMG are dual to a two-dimensional CFT.  What we have seen is that in a three-dimensional theory of gravity coupled to matter fields, one may be able to find a second Virasoro when the gauge fiber is treated geometrically.  This is similar in spirit to the study of an $\textit{AdS}_2$ background with gauge field \cite{Hartman:2008dq}, as well as to the proposed holographic description of extremal Reissner-Norstrom black holes in \cite{Hartman:2008pb}.

\subsection{Supersymmetry of the Solutions}

Here we will deal with the type IIA and heterotic cases; type IIB is similar to both this section as well as to \cite{Compere:2008cw}.  The metric is as in (\ref{10dBG-2}) with $S^3$ radius $\sqrt{k}$ while $H_3$ takes the form
\be
H_3 = 2 \cosh \!\varrho \, \sinh \!\varrho \, \big(   k \di t\wedge \di\!\varrho\wedge \di \phi - \textsc{h} \sqrt{k} \di\!\varrho \wedge \di\varphi \wedge ( \di\phi + \sigma \di t ) \big) + 2 k \textrm{Vol}(S^3)\, , 
\ee
where $\textrm{Vol}(S^3)$ is the volume form on the unit 3-sphere and $\sigma=\pm 1$ determines whether we deform the holomorphic or antiholomorphic side of the worldsheet CFT.  In particular, in a suitably chosen frame this is
\be
H_3 = \frac{2}{\sqrt{k}} e^{\hat{t}} \wedge e^{\hat{\varrho}} \wedge e^{\hat{\phi}} + 2 k \textrm{Vol}(S^3) \, .
\ee
The Riemann tensors with $\pm \tfrac{1}{2}H_3$ torsion both vanish, so the integrability conditions on the gravitini variations are always satisfied.  This locally guarantees the existence of a solution to the gravitini variations but not globally, as we will see.  The dilatino variation constrains the Killing spinor $\varepsilon$ to satisfy the projection
\be
\Gamma^{\hat{t}\hat{\varrho}\hat{\phi}\hat{\alpha_1}\hat{\alpha_2}\hat{\alpha_3}} \varepsilon = - \varepsilon \, ,
\ee
which poses no trouble.

We still have to solve for the Killing spinor which must satisfy $\nabla_\mu^{(\beta)} \varepsilon_\beta=0$, where $\nabla_\mu^{(\beta)}$ refers to the connection with torsion $\tfrac{1}{2} \beta H_3$ and we have split $\varepsilon$ into Majorana-Weyl components, of which there are two for type IIA ($\beta=\pm$) and only one for heterotic ($\beta=-$).  Since a solution is guaranteed locally, the only concern we have is whether the solution of the Killing spinor equation respects the periodicity of the fibered circle.  To check this, it turns out that we only need to solve two of the components of the gravitino variation:
\begin{subequations}
\bea
\label{eqn:killing-rho}
\nabla^{(\beta)}_\varrho \varepsilon_\beta & =& \Big( \p_\varrho - \frac{\beta}{2} \Gamma^{\hat{t}\hat{\phi}} \Big) \varepsilon_\beta = 0  \, , \\
\label{eqn:killing-phi}
\nabla^{(\beta)}_\varphi \varepsilon_\beta &=& \Big( \p_\varphi - \frac{\textsc{h}}{2\sqrt{k}} (1+ \beta\sigma) \big( \sinh \!\varrho \Gamma^{\hat{t}\hat{\varrho}} + \sigma \cosh\!\varrho \Gamma^{\hat{\phi}\hat{\varrho}} \big) \Big) \varepsilon_\beta = 0 \, .
\eea
\end{subequations}
The first thing to notice is that when $\beta\sigma = -1$, \re{eqn:killing-phi} implies that $\varepsilon$ is independent of $\varphi$.  This will always yield a space-time supersymmetry: in the type IIA case, this means that we will always have at least one globally defined Killing spinor, $\varepsilon_{-\sigma}$; in the heterotic case, where $\beta=-$, this means that we will have a Killing spinor for all values of the moduli when $\sigma=1$, so this corresponds to deforming the bosonic side of the heterotic worldsheet.

On the other hand, when $\beta\sigma=+1$, the first equation \re{eqn:killing-rho} is satisfied by
\be
\varepsilon_\beta = \Big( \cosh\tfrac{\varrho}{2} + \beta\sinh\tfrac{\varrho}{2} \, \Gamma^{\hat{t}\hat{\phi}} \Big) \eta_\beta \, ,
\ee
which we can then use to simplify \re{eqn:killing-phi}:
\be
\label{eqn:deformed-susy}
\Big( \cosh\tfrac{\varrho}{2} + \beta\sinh\tfrac{\varrho}{2} \, \Gamma^{\hat{t}\hat{\phi}} \Big) \Big(   \p_\varphi - \beta\frac{ \textsc{h}}{\sqrt{k}} \Gamma^{\hat{\phi}\hat{\varrho}}      \Big)  \eta_\beta = 0 \, .
\ee
The solution to \re{eqn:deformed-susy} will depend on $\cos\!\big(\frac{\textsc{h}}{\sqrt{k}}\varphi\big)$, and since our Killing spinors should be periodic or antiperiodic around the fibered circle, this Killing spinor will only be globally defined when
\be
\label{eqn:enhanced-susy}
\frac{2\textsc{h} R}{\sqrt{k}} \in \mathbb{Z} \, .
\ee
In the type IIA case, this means that half of undeformed supersymmetries will be broken when (\ref{eqn:enhanced-susy}) is satisfied, in which case we will have all the supersymmetry of the undeformed case.  In the heterotic case, since we are fixed to have $\beta=-$, this means that the deformation corresponding to $\sigma=-1$ will generically break all of supersymmetry except when (\ref{eqn:enhanced-susy}) holds, in which case it will break no supersymmetry and, hence, corresponds to deforming the supersymmetric side of the heterotic worldsheet.  This is in complete agreement with the worldsheet analysis performed in section \ref{sec:space-time-susy}.

The type IIB theory differs from type IIA only in the GSO projection, so from the worldsheet arguments of section \ref{sec:space-time-susy} we expect the same conclusion to hold for type IIB.  This is in agreement with the G\"odel space-time analysis of \cite{Compere:2008cw}.


\section{Discussion}
\label{sec:discussion}

As we have seen, when we consider space-time backgrounds that consist of an $S^1$ bundle over \wads, we find that we can tune the radius $R$ of the $S^1$ so that when $\frac{2\textsc{h}R}{\sqrt{k}}$ takes rational values $\frac{\mu}{\rho}$, we obtain two commuting space-time Virasoro algebras with central charges (taking, for instance,  $w>0$)
$$
c_{\mathrm{st}} = 6k\left( \rho w + \mu \omega \right)\, , \qquad  \bar{c}_{\mathrm{st}}=6kw \, ,
$$
whereas when it is irrational we obtain only one space-time Virasoro algebra together with a commuting global charge, 
$L_0^{\mathrm{st}}$.  Generically, half of space-time supersymmetry is broken in the type II case and all of space-time supersymmetry is broken in the heterotic case (unless we deform the bosonic side, in which case none is broken) except when $\frac{2\textsc{h}R}{\sqrt{k}}$ is integer ({\it e.g.}, $\rho=1$), in which case no space-time supersymmetries are broken.  Dual CFTs with differing left and right central charges usually indicate a diffeomorphism-violating term in the bulk lagrangian, so the cases where $\rho \neq 1$ are rather puzzling because these backgrounds arise as solutions to a consistent truncation of supergravity that has no diffeomorphism-violating terms~(see section~\ref{sec:kk-reduction}).

We can always realize the deformed worldsheet sigma model as a $\mathbb{Z}$-orbifold of an $\textit{AdS}_3 \times \bb{R}$ WZW model with discrete torsion phases.  It is tempting to conjecture that the dual QFT is also an orbifold of the CFT dual to $\textit{AdS}_3 \times \bb{R}$, {\it e.g.} by an operator of the form $\exp\!\left(2\pi i \frac{2\textsc{h}R}{\sqrt{k}}L_0^{\mathrm{st}}\right)$, but this appears to be incorrect in part because it does not reproduce the term in the central charge related to winding around the $S^1$, $\omega$.  Similarly, in searching for the brane construction it would be tempting to guess that it is an orbifold of the full F1-NS5 system, but this is also evidently untrue since the $\bb{Z}$ by which we orbifold is embedded in a $U(1)$ isometry of the near-horizon geometry that does not extend to an isometry of the full geometry.

Even at the points of enhancement where we expect two Virasoro algebras, the interpretation of the dual theory appears slightly murky.  To see this, let's focus on the supersymmetric models with $\rho=1$ which have the property that the left and right 
central charges of the dual CFT are equal, at least in the sector with no winding along the circle (corresponding to the supergravity limit where winding modes decouple).  We can then formally define the dual two-dimensional space-time CFT, coordinatized by $(x,\bar x)$, by declaring the two sets of space-time charges $\{L_n^\mathrm{st} \}$ and $\{\bar{L}_{m}^\mathrm{st} \}$ to be the Laurent modes of a 
conserved  stress-energy tensor with components $T^\text{st} (x)$ and $\bar{T}^\text{st} (\bar x)$.   We would then like to interpret worldsheet physical states, with quantum numbers $(j,m,\bar m,p)$, as modes of primary operators in the dual theory; using a change of basis and temporarily restricting to unflowed representations, $w=0$, suggests
\be
\Phi_{j,p} (x,\bar x) = \sum_{m,\bar m} x^{-j-m} \bar{x}^{-j-\bar m} \Phi_{j\, m\, \bar m,p}\, ,
\ee
where the sum over $m$ and $\bar m$ should be unrestricted for $\Phi_{j,p}(x,\bar{x})$ to transform in a representation of the dual conformal algebra. However, solving the on-shell condition on the worldsheet 
\be
\Delta_{\textrm{ws}} = -\frac{j(j-1)}{k-2} + \frac{\cos^2 \alpha}{2 R^2} \left(p-\mu m \right)^2 
+ N +\Delta_\text{other} = 0
\ee
will generically force $m$ to take specific values, implying that $\Phi_{j,p}(x,\bar{x})$ will not transform properly.
This conclusion is avoided if one tunes the $ S^1$ momentum to take the value $p = \mu m$, but the implication is the same: it appears that only a subspace of the dual theory will form representations of the dual conformal algebra.  The same argument obtains for spectrally flowed operators, $w\neq 0$.

Given these challenges, we leave the identifications of the dual QFT and the brane construction for future work.  Ultimately, our goal is to understand the QFTs dual to space-like stretched \wads~and null \wads, which have black holes and no CTCs, but this is currently beyond the scope of the methods used here (representations of $\widehat{\SL}$ in the parabolic and hyperbolic basis are harder to handle).  Until the day that we can extend these methods to space-like \wads~arrives, we content ourselves with squeezing every drop out of time-like \wads.

\section*{Acknowledgements}
We would like to thank R. Argurio, Ch.Chaplin, G. Giribet, D. Orlando, M. Rangamani, A. Maloney, G. Comp\`ere, D. Anninos, A. Strominger, J. Polchinski, D. Berenstein, T.Creutzig, P. Ronne, and  J. Troost for various fruitful discussions and exchanges.  The works of S.D. are funded by the European Commission though the grant PIOF-GA-2008-219950  (Home institution: Universit\'e Libre de Bruxelles, Service de Physique Th\'eorique et Math\'ematique, Campus de la Plaine, B-1050 Bruxelles, Belgium).  The work of D.I. was supported in part by the Agence Nationale de la Recherche grant "STR-COSMO" (ANR-09-BLAN-0157).  The work of S.D., J.L., and M.R. was supported in part by the National Science Foundation under Grant No. PHY05-51164 and Grant No. PHY07-57035.


\appendix

\section{Parafermions and $\widehat{SL(2,\bb{R})}$}
\label{app:parafermions}

Parafermions were originally studied by Zamolodchikov and Fateev in the context of the two-dimensional Ising model by generalizing the $\mathbb{Z}_{2}$ symmetry to $\mathbb{Z}_{k}$ \cite{Fateev:1985mm}. 
As in the $\mathbb{Z}_{2}$ case, the $\mathbb{Z}_{k}$ model is expected to have a critical point where the $\mathbb{Z}_{k}$ symmetry is realized and is described by fields $\phi_{(m,\mb)}$ with $\mathbb{Z}_{k}\times \mathbb{Z}_{k}$ charge $\left(\frac{1}{2}(m+\mb),\frac{1}{2}(m-\mb)\right)$.  When $k=2$, these fields are free fermions while in the general case, they are promoted to parafermions $\psi_{m}\sim\phi_{(m,0)}$, $\bar{\psi}_{\mb}\sim\phi_{(0,\mb)}$ --- these describe the $SU(2)/U(1)$ coset model.  The generalization to $SL(2,\mathbb{R})$ was studied in \cite{Lykken:1988ut} and the construction of primaries is similar to that of $SU(2)$, the main difference being that there is no known, underlying, Ising-type model.  The relevant OPEs of the holomorphic parafermions are
\begin{eqnarray}
\label{parafopes}
\psi_{m}(z)\psi_{m'}(z') &=& c_{m,m'}(z-z')^{\Delta_{m+m'}-\Delta_{m}-\Delta_{m'}}\Big(\psi_{m'+m}(z')+O((z-z'))\Big) \, , \nonumber\\
\psi_{m}(z)\psi_{m}^{\dag}(z')&=&(z-z')^{-2\Delta_{m}}\Big(I+\frac{2\Delta_{m}}{c_{p}}T_{p}(z')(z-z')^{2}+O((z-z')^{3})\Big) \, ,
\end{eqnarray}
with the parafermion central charge given by
\begin{eqnarray}\label{parafcentral}
c_{p} = \frac{3k}{k-2} - 1 \, .
\end{eqnarray}
The complete Fock space is constructed by acting with $\psi_1$ and $\psi_1^\dag$ on the parafermion principal fields $\Phi^{j}_{m}$ \cite{Lykken:1988ut}. So, the complete space of fields can be denoted
\begin{eqnarray}
\mathcal{H}=\bigoplus_{l=0}^{n-1}\left[\Phi^{l,\bar{l}}_{(l,\bar{l})}\right]
\end{eqnarray}
Where $\left[\Phi^{l,\bar{l}}_{(l,\bar{l})}\right]$ correspond to the conformal block generated by all the descendants of $\Phi^{l,\bar{l}}_{(l,\bar{l})}$.

\subsection{Relation to $\widehat{SL(2,\bb{R})}$ WZW Model}

Consider a free time-like boson $T(z,\bar{z})=T(z)+\bar{T}(\bar{z})$ with OPE $T(z,\bar{z})T(0,0)\sim \ln|z|^2$.
Then we can take the direct product the free boson and the $\mathbb{Z}_{k}$ parafermion system in order to write the $\widehat{SL(2,\bb{R})}_{k}$ currents as
\begin{eqnarray}
J^{+}&=&i\sqrt{k}\,\psi_{1}\,e^{-i\sqrt{\frac{2}{k}}T}(z)\, , \nonumber\\
J^{-}&=&-i\sqrt{k}\,\psi_{1}^{\dag}\,e^{i\sqrt{\frac{2}{k}}T} (z) \, ,\nonumber\\
J^{3}&=&i\sqrt{\tfrac{k}{2}}\,\partial T(z).
\end{eqnarray}
The wieghts of the parafermionic currents are $\Delta(\psi_{1})=\Delta(\psi_{1}^{\dag})=1+\frac{1}{k}$ while the $SL(2,\bb{R})$ currents have OPEs
\begin{eqnarray}
J^3(z) J^3(0) \sim -\frac{k}{2z^2} \, ,  \qquad  J^{+}(z)J^{-}(0)\sim\frac{k}{z^{2}}-2\frac{J^{3}(0)}{z}\, ,  \qquad J^{3}(z)J^{\pm}(0)\sim\pm\frac{J^{\pm}(0)}{z} \, ,
\end{eqnarray}
and similarly for the anti-holomorphic currents.  The WZW primaries are then given by
\begin{eqnarray}
\mathcal{V}_{m,\mb}^{l,\bar{l}}=\Phi^{l,\bar{l}}_{(m,\mb)} e^{-i\sqrt{\frac{2}{k}}mT- i\sqrt{\frac{2}{k}}\mb\bar{T}} \, .
\end{eqnarray}

\subsection{Free Field Representations}

A free field representation for the $SL(2,\bb{R})$ parefermions can be constructed in terms of two free bosons (for the $SU(2)$ case, see \cite{Gerasimov:1989mz}),
$\chi$ and $s$, which have the OPEs $\chi(z)\chi(0) \sim -\ln(z)$, $s(z)s(0)\sim -\ln(z)$.
Then we can write
\begin{eqnarray}
\label{freeparafrep}
\psi_{1}(z)&=&\frac{1}{\sqrt{2}} \Big(\partial\chi+i\sqrt{\frac{k-2}{k}}\partial s\Big)e^{i\sqrt{\frac{2}{k}}\chi} \, , \nonumber\\
\psi_{1}^{\dag}(z)&=&-\frac{1}{\sqrt{2}} \Big(\partial\chi-i\sqrt{\frac{k-2}{k}}\partial s\Big)e^{-i\sqrt{\frac{2}{k}}\chi} \, , \nonumber\\
T_p(z)&=&-\frac{1}{2} \partial\chi \p\chi-\frac{1}{2}\partial s\p s+\frac{1}{\sqrt{2(k-2)}}\partial^{2}s \, .
\end{eqnarray}
In this representation, the parafermion principal fields are given by
\begin{eqnarray}
\Phi^{j}_{m}
=e^{i\sqrt{\frac{2}{k}}m\chi+j\sqrt{\frac{2}{k-2}}s} \, ,  \qquad \Delta(\Phi^{j}_{m})=-\frac{j(j-1)}{k-2}+\frac{m^{2}}{k} \, ,
\end{eqnarray}
and so we have
\begin{eqnarray}
\psi_{1}(z)\Phi^{j}_{m}(w)&\sim&(z-w)^{\frac{2m}{k}}\left(-\frac{i(m+j)}{\sqrt{k}(z-w)}\Phi^{j}_{m+1}+O(1)\right) \, , \nonumber\\
\psi_{1}^{\dag}(z)\Phi^{j}_{m}(w)&\sim&(z-w)^{-\frac{2m}{k}}\left(\frac{i(m-j)}{\sqrt{k}(z-w)}\Phi^{j}_{m-1}+O(1)\right) \, .
\end{eqnarray}


\section{Current-Current Deformations and Lattice Rotations}
\label{app:latticerotations}

A WZW model is characterized by its left and right currents which each satisfy an affine Kac-Moody algebra $\hat{\mathfrak{g}}_{k}\times\hat{\bar{\mathfrak{g}}}_{k}$. Generically, the Hilbert space will be decomposed as
\begin{eqnarray}
\mathcal{H}=\bigoplus_{(Q,\overline{Q})\in \Lambda} \mathcal{H}_{Q}\otimes\overline{\mathcal{H}}_{\overline{Q}}
\end{eqnarray}
where $(Q,\overline{Q})$ are elements in the lattice of charges $\Lambda$. $\Lambda$ is characterized by the eigenvalues of the generators of the Cartan subalgebra $\mathfrak{h}\times\bar{\mathfrak{h}}\subseteq \hat{\mathfrak{g}}_{k}\times\hat{\bar{\mathfrak{g}}}_{k}$.  It was shown in \cite{Forste:2003km} that a marginal deformation of the form $c_{ij}J^{i}\overline{J}^{j}$, with $(J,\overline{J})\in \mathfrak{h}\times\bar{\mathfrak{h}}$, can be implemented by an $\frac{O(d,\bar{d})}{O(d)\times O(\bar{d})\times O(d,\bar{d};\mathbb{Z})}$ rotation on the charge lattice of the undeformed model. To implement this, we write the WZW model as a coset model times a toroidal CFT with an orbifold action: $\hat{\mathfrak{g}}_{k}\simeq \left(\hat{\mathfrak{g}}_{k}/\hat{\mathfrak{h}}\otimes T_{\Lambda}\right)/\Gamma$.
The deformed toroidal sigma model then has the general form
\begin{equation}
S = \frac{1}{2\pi \alpha'} \int {\rm d}^2 z \, (G_{ij} + \Delta G_{ij} + B_{ij} + \Delta B_{ij})\partial X^i \bar \partial X^j \, ,
\end{equation}
with a constant metric $G+\Delta G$ and $B$-field $B+\Delta B$ that depend on the parameters of the deformation through $\Delta G$ and $\Delta B$. The fields are subject to the toroidal identifications $X^j\sim  X^j +2\pi R^j_{i}$ for all $i,j=1,\ldots,\dim(\mathfrak{h})$.

The dimensions of operators can be analyzed in the usual way.  Calling the deformed fields $\hat{G}$ and $\hat{B}$, we can write the dimensions of operators in the deformed theory as
\be
\hat{\Delta} = \frac{\alpha'}{4} \hat{G}^{ij} \hat{p}_{L,i} \hat{p}_{L,j} \qquad  \textrm{and} \qquad  \hat{\bar{\Delta}} = \frac{\alpha'}{4} \hat{G}^{ij} \hat{p}_{R,i} \hat{p}_{R,j} 
\ee
where, as usual,
\be
\hat{p}_{\stackrel{L}{\textrm{\tiny{\it R}}}}{}_{,i} = (R^{-1})^j_i n_j +  \tfrac{1}{\alpha'}\Big( \hat{B}_{ij} \pm \hat{G}_{ij} \Big) R^j_k w^k
\ee
for $n_i, w^j \in \bb{Z}$.  What was shown in \cite{Forste:2003km} was that we can also write these dimensions as
\be
\hat{\Delta} = \frac{\alpha'}{4} G^{ij} p'_{L,i} p'_{L,j} \qquad  \textrm{and} \qquad  \hat{\bar{\Delta}} = \frac{\alpha'}{4} G^{ij} p'_{R,i} p'_{R,j} \, ,
\ee
where we contract with the \emph{undeformed} metric $G$, the momenta $p'_{L,R}$ are determined by
\be
\left( \begin{array}{c} p'_{L} \\ p'_{R} \end{array} \right)  =  \Omega \left( \begin{array}{c} p_L \\ p_R \end{array}\right) \, ,  \qquad  \Omega \in \frac{O(d,\bar{d})}{O(d)\times O(\bar{d})\times O(d,\bar{d};\mathbb{Z})} \, ,
\ee
and $p_{L,R}$ are the momenta of the \emph{undeformed} theory.  Thus, we have a way of writing down the spectrum of the deformed theory directly from that of the undeformed theory.

\section{BRST Formalism for Coset Models}
\label{app:brst-coset}

One way to realize a level-$k$ coset model $(G/H)_k$, $H\subseteq G$, is as a product
\be
\left(G_k \times H_{-k-2 h(H)} \times \textit{ghosts} \right) / \sim \, ,
\ee
where the equivalence relation $\sim$ is defined by a suitable BRST operator and $h(H)$ is the dual Coxeter number of $H$ \cite{Karabali:1989dk,Hwang:1993nc}. 
We can also realize the coset as a gauged WZW model, and the $bc$ ghost system here is precisely that which is required to gauge the subgroup $H\subseteq G$.  The ghost system has central charge $-2d_H$, where $d_H=\dim(H)$, so that the total central charge is given by
\be
c_{G/H} =c_{G,k}+c_{H,-k-2h(H)}-2d_H = \frac{k\, d_G}{k+h(G)} - \frac{k\, d_H}{k+h(H)} = c_{G,k} - c_{H,k}  \, .
\ee
The BRST operator is given by
\be
\label{eqn:brst-coset}
Q_{G/H}=\sum_{i=1}^{d_H}\oint\frac{\di z}{2\pi i}\, :c^{i}\big(J_G^{i}+\widehat{J}_H^{i}+\tfrac{1}{2}J^{i}_{gh}\big): \, ,
\ee
where $J_{gh}^i \sim i f(H)^{ij}_{\ph{ij}k}\! : \!\!b_j c^k\!\!:$.  As usual, physical states are defined by the cohomology of $Q_{G/H}$, $H_{Q_{G/H}}\left( \mathcal{H}_G \otimes \mathcal{H}_H \otimes \mathcal{H}_{gh} \right)$.  Primary states are in the ghost vacuum --- $c_{n\geq 1}^i |0\rangle_{gh} = 0 = b_{n\geq 0}^i |0\rangle_{gh}$ --- and, therefore, annihilated by $\oint \!: \!\!c^i J^i_{gh}\!\! :$.  Thus, the physical state condition for primary operators simply sets 
\be
\big( J^i_G + \widehat{J}_H^i \big) |\textit{primary}\rangle = 0 \, .
\ee

The extension to supersymmetric models is straightforward: $(G/H)_k$ is realized as $\big(G_k \times H_{-k} \times \textit{superghosts}\big) / \sim$, where the equivalence relation is now defined by a BRST operator constructed out of the `total' $G$ and $H$ currents $\mathcal{J}^i \sim J^i + i f^i_{\ph{i}jk} \psi^j \psi^k$, {\it et cetera}.  The ghost central charge in this case is $-3d_H$ so that the total central charge is
\be
c_{G/H,k} = c_{G,k} + c_{H,-k} - 3 d_H = \frac{(k+h(G))\, d_G}{k} - \frac{(k+h(H))\, d_H}{k} + \frac{(d_G - d_H)}{2} = c_{G,k} - c_{H,k} \, ,
\ee
as it should be.


\section{space-time Virasoro from the BRST Formalism}
\label{app:bosonicbrst}

\subsection{Undeformed Case}

Following appendix \ref{app:brst-coset}, we realize $SL(2,\bb{R})_{k}/U(1)$ as $\big(SL(2,\bb{R})_k \times U(1) \times bc \big)/\sim$, where $b$ has weight $1$, $c$ has weight $0$, and $\sim$ is defined through the BRST operator $Q_{G/H}$ in (\ref{eqn:brst-coset}).  The $U(1)$ of the BRST formulation must have level opposite that of the $U(1)$ of the coset.  In particular for the time-like deformation, the $U(1)$ of the coset, generated by $K = i\sqrt{\frac{k}{2}} \,\p T$,\footnote{This is not to be confused with the $J^3_G$ that will appear below, which is the $J^3_G$ of the $SL(2,\bb{R})$ that appears in the BRST formulation.} has the level $-\frac{k}{2}$ so that the level of the $U(1)$ in the BRST formulation must be $+\frac{k}{2}$ --- let's denote this as $\widehat{J}_H = i\sqrt{\frac{k}{2}}\, \p Y$, where $Y$ is a space-like boson.  Since $H=U(1)$ is abelian, $J_{gh}=0$ and the BRST operator for the coset theory (not to be confused with the BRST operator for the full worldsheet theory, which we'll call $Q_B$) is simply
\be
Q_{G/H} \equiv \oint \frac{\di z}{2\pi i} : \!c \big( J_G^3 + \widehat{J}_H \big) \!: \, .
\ee
A primary state in the BRST formulation is annihilated by $c_{n>0},~b_{n\geq 0},~J^a_{G,n>0},$ and $\widehat{J}_{H,n>0}$.  Thus, a \emph{physical} primary state must have equal and opposite eigenvalues of $J^3_{G,0}$ and $\widehat{J}_{H,0}$.

Just to be pedantic, we are writing the Hilbert $\mathcal{H}$ space of our original $SL(2,\bb{R})_k$ as
\bea
\mathcal{H}\big(SL(2,\bb{R})_k\big)   &\cong&   \mathcal{H}\left( \frac{\big(SL(2,\bb{R})/U(1)\big)_k \times U(1)_T}{\bb{Z}} \right)   \nonumber \\  
&\cong&  H_{Q_{G/H}} \left( \mathcal{H}\left( \frac{SL(2,\bb{R})_k \times U(1)_Y \times bc \times U(1)_T}{\bb{Z}} \right)\right) \, ,
\eea
where the $\bb{Z}$ acts to match the eigenvalue of $J_{G,0}^3$ with that of $K_0$ (in sectors with zero spectral flow).  Below, we'll restrict to states with $j=\bar{m}=0$ in order to work with holomorphic operators, so we'll suppress these subscripts for convenience and will write a full $SL(2,\bb{R})$ primary as ${\cal V}^G_m e^{-i\sqrt{\frac{2}{k}} m ( t + \phi )}$, where $\V^G_m$ refers to a primary of the $SL(2,\bb{R})$ of the BRST formulation.

Given a WZW primary state $|m\rangle$ which has $L_0 = 0$, we can generate a state with dimension $1$ (and the same ghost number) by acting with $c_0 b_{-1}$, $J^a_{G,-1}$, $\widehat{J}_{H,-1}$, or $K_{-1}$, however up to a $Q_{G/H}$-exact piece, acting with $J^3_{G,-1}$ is the same as acting with $\widehat{J}_{H,-1}$ and so we will ignore it.  Furthermore, $c_0 b_{-1}$ will not lead to a $Q_{G/H}$-closed state because $\{ Q_{G/H}, c_0 b_{-1}\} \sim c_0 \big( J^3_{G,-1} + \widehat{J}_{H,-1} \big)$, which cannot cancel with $Q_{G/H}$ acting on the other states.  So, our most general guess for space-time Virasoro charges is
\be
L^{\mathrm{st}}_m = \oint \frac{dz}{2\pi i} \bigg\{  a_m i\sqrt{\frac{k}{2}}\, \p T {\cal V}^G_m + b^+_m : J_G^+ {\cal V}^G_{m-1} : + b^-_m :J_G^- {\cal V}^G_{m+1}: + c_m i\sqrt{\frac{k}{2}}\, \p Y \, {\cal V}^G_{m} \bigg\} e^{-i\sqrt{\frac{2}{k}} m ( T + Y )} \, .
\ee

We have two constraints to impose: first, the state must be $Q_{G/H}$-closed; second, the state must be $Q_B$-closed, which means it must be annihilated by $L_1$.  These imply the two conditions
\bea
\label{eqn:physical-states}
(m-1) b_m^+ - (m+1) b_m^- + \frac{k}{2} c_m &=& 0  \, , \\
m (a_m - c_m) + (m-1) b_m^+ + (m+1) b_m^- &=& 0 \, .
\eea
For reference, we include some relevant OPEs:
\begin{subequations}
\bea
\V^G_m(z_1)\V^G_n(z_2) &\sim & \V^G_{m+n}(z_2) \, , \\
\V^G_m(z_1) : J_G^\pm \V^G_n : (z_2) &\sim & -\frac{m}{z_{12}} \V^G_{m+n\pm 1}(z_2)  \, , \\
%
%
%
%
:J^\pm_G\V^G_n:(z_1) :J_G^\pm\V^G_m:(z_2) &\sim  & -\frac{mn}{(z_{12})^2} \V^G_{m+n\pm 2}(z_2) - \frac{mn}{z_{12}} :\V^G_{m\pm 1} \p\V^G_{n\pm 1}:(z_2)  \nonumber \\
&& + \frac{(m-n)}{z_{12}} :J_G^\pm\V^G_{m+n\pm 1}:(z_2)  \, ,   \\
:J^\pm_G\V^G_n:(z_1) :J^\mp_G \V^G_m:(z_2) &\sim& \frac{(k-mn\pm 2(n-m))}{(z_{12})^2} \V^G_{m+n}(z_2) + \frac{1}{z_{12}} \Big[ (k\pm 2(n-m)):\V^G_m \p\V^G_n:  \nonumber \\
&& - \,mn \!:\!\V^G_{m\pm 1} \p \V^G_{n\mp 1} \!:\! \mp \,2\!:\!J^3_G\V^G_{m+n}: - \,n\!:\!J_G^\pm \V^G_{m+n\mp 1}\!:  \nonumber \\
&& + \, m \!:\!J_G^\mp\V^G_{m+n\pm 1}\!:\! \Big](z_2) \, .
%
\eea
\end{subequations}
Some of these can be simplified bit by noting that $:\p \V^G_m: = m:\V^G_{m-1} \p \V^G_1:$, and we'll make use of this below.  Define $L^{\mathrm{st}}_m \equiv \oint \frac{\di z}{2\pi i} {\cal L}_m(z)$, then
\bea
\oint_{C_{z_2}}\frac{\di z_1}{2\pi i} {\cal L}_n(z_1){\cal L}_m(z_2) &\sim & \Bigg\{ \sum_{s=\pm} b_n^s b_m^s \bigg[ \big( mn + 1 - s(m+n) \big) in\sqrt{\tfrac{2}{k}} \big( \p Y + \p T\big)  \V^G_{m+n}  \nonumber \\
&& \!\!\!\!\!\! + (m-n) :J_G^s \V^G_{m+n-s}:   - \big( mn + 1 - s(m+n)\big)n \V^G_{m+n-1} \p\V^G_1 \bigg]   \nonumber  \\
&& \!\!\!\!\!\!\!\!\!\!\!\!\!\! + \sum_{s=\pm} b_n^s b_m^{-s} \bigg[ -\big(k-3-mn+s(n-m)\big) in\sqrt{\tfrac{2}{k}} \big( \p Y + \p T \big) \V^G_{m+n}  \nonumber \\
&& \!\!\!\!\!\! + \big( nk - 3n - mn^2 + sn(m+n) - s(k-2)\big) \V^G_{m+n-1} \p\V^G_1 - 2s:J_G^3\V^G_{m+n}: \nonumber  \\
&& \!\!\!\!\!\! - (n-s) :J_G^s \V^G_{m+n-s}: + (m+s) :J_G^{-s}\V^G_{m+n+s}: \bigg]   \nonumber \\
&& \!\!\!\!\!\!\!\!\!\!\!\!\!\! + (a_n - c_n) \sum_{s=\pm} b_m^{s} \bigg[ i mn^2\sqrt{\tfrac{2}{k}} \big(\p Y + \p T \big) \V^G_{m+n} + m:J_G^s \V^G_{m+n-s}:  \nonumber \\
&& \!\!\!\!\!\!  - mn(n+s) \V^G_{m+n-1} \p\V^G_1 - in\sqrt{\tfrac{k}{2}}\p T\, \V^G_{m+n} \bigg]   \nonumber \\
&& \!\!\!\!\!\!\!\!\!\!\!\!\!\! + (a_m-c_m) \sum_{s=\pm} b_n^s  \bigg[ i mn^2\sqrt{\tfrac{2}{k}} \big(\p Y + \p T\big) \V^G_{m+n} - n:J_G^s \V^G_{m+n-s}:  \nonumber \\
&& \!\!\!\!\!\!  - mn(n-s) \V^G_{m+n-1} \p\V^G_1 + im\sqrt{\tfrac{k}{2}}\p T\, \V^G_{m+n} \bigg]   \nonumber \\
&&\!\!\!\!\!\!\!\!\!\!\!\!\!\! + (a_n-c_n)(a_m-c_m) \bigg[ \big(-\tfrac{kn}{2}-mn^2\big) \V^G_{m+n-1}\p\V^G_1    +  i\sqrt{\tfrac{k}{2}}\big( n\p Y + m\p T\big) \V^G_{m+n} \nonumber \\
&& \!\!\!\!\!\! + imn^2\sqrt{\tfrac{2}{k}} \big( \p Y + \p T \big) \V^G_{m+n} \bigg]   \nonumber \\
&& \!\!\!\!\!\!\!\!\!\!\!\!\!\! + (a_n-c_n) c_m \bigg[ i\sqrt{\tfrac{k}{2}}(m+n)\big( \p Y + \p T\big) \V^G_{m+n} - \tfrac{kn}{2} \V^G_{m+n-1}\p\V^G_1 \bigg]  \nonumber \\
&& \!\!\!\!\!\!\!\!\!\!\!\!\!\! + (a_m-c_m) c_n \bigg[ -\tfrac{kn}{2}\V^G_{m+n-1}\p\V^G_1 \bigg]   \nonumber \\
&& \!\!\!\!\!\!\!\!\!\!\!\!\!\! + \sum_{s=\pm} \big( mc_m b_n^s - n c_n b_m^s \big) i\sqrt{\tfrac{k}{2}}\big(\p Y + \p T\big) \V^G_{m+n} \Bigg\} e^{-i\sqrt{\frac{2}{k}} (m+n) (Y+T)}(z_2) \, .
\eea
Collecting the coefficient of $:J_G^s \V^G_{m+n-s}:$ and demanding it equal $(n-m)b_{m+n}^s$ in order it satisfy the Virasoro algebra yields
\be
(m-n)b^s_n b^s_m - (n-s)b_n^s b_m^{-s} + (m-s) b_n^{-s} b_m^s + m(a_n - c_n)b_m^s - n (a_m - c_m) b_n^s = (n-m)b_{m+n}^s \, .
\ee
If we first restrict ourselves to cases where $m,n\neq 0$, and $m+n\neq 0$, we can multiply this equation by $mn$ and simplify by using the physical state condition (\ref{eqn:physical-states})
\be
s(m+n) \Big[ (m-n) b_n^s b_m^s + n b_n^s b_m^{-s} - m b_n^{-s} b_m^s \Big] = mn(n-m) b_{m+n}^s \, .
\ee
We can solve this constraint by setting
\be
b_m^s = - \frac{sm}{2} \, .
\ee
and input this into the physical state conditions (\ref{eqn:physical-states}) to find
\be
a_m - c_m = -1  \qquad  \textrm{and}  \qquad  c_m = \frac{2m^2}{k} \, .
\ee

Working with this solution drastically simplifies the OPE to
\bea
\oint_{C_{z_2}}\frac{\di z_1}{2\pi i} {\cal L}_n(z_1){\cal L}_m(z_2) &\sim &   \Bigg\{  \sum_{s=\pm} (n-m) b_{n+m}^s  :J_G^s \V^G_{m+n-s}:      \nonumber \\
&& + \Big( \tfrac{mn}{2}\big( -nk   \big)   -\tfrac{kn}{2} + mn^2  + n^3 \Big)  \V^G_{m+n-1} \p\V^G_1   \nonumber \\
&&    + \Big( \tfrac{mn}{2}(nk) + \tfrac{k}{2}n  - m^3 - m^2 n   \Big) i \sqrt{\tfrac{2}{k}} \p Y\,\V^G_{m+n} \nonumber \\
&&    + \Big( \tfrac{mn}{2}(kn)+\tfrac{k}{2}m - m^3 - m^2 n\Big) i \sqrt{\tfrac{2}{k}} \p T\, \V^G_{m+n}     \Bigg\} e^{-i\sqrt{\frac{2}{k}} (m+n) (Y+T)}(z_2) \, . \nonumber
\eea
Now we deal with two cases separately.  First if $m+n\neq 0$, then we can write $\V_{m+n-1}\p\V_1 = \frac{1}{m+n} \p\V_{m+n}$.  When we compute $[L^{\mathrm{st}}_m,L^{\mathrm{st}}_n]$, we will also integrate over $z_2$ and can, therefore, integrate by parts, which amounts to making the replacement
\be
\V^G_{m+n-1}\p\V^G_1  \longrightarrow i\sqrt{\tfrac{2}{k}}(\p Y + \p T) \, .
\ee
In this case, we find
\bea
[L^{\mathrm{st}}_n, L^{\mathrm{st}}_m] &= & (n-m) \oint \frac{\di z}{2\pi i}   \Bigg\{  \sum_{s=\pm} b_{n+m}^s  :J_G^s \V^G_{m+n-s}:   + c_{n+m} i \sqrt{\tfrac{k}{2}} \p Y \,\V^G_{m+n} \nonumber \\
&&   \qquad \qquad\qquad\qquad + a_{n+m} i \sqrt{\tfrac{k}{2}} \p T\, \V^G_{m+n}     \Bigg\} e^{-i\sqrt{\frac{2}{k}} (m+n) (Y+T)}(z)   \nonumber \\
&=& (n-m)L_{n+m} \, ,  \qquad \textrm{for} \qquad m+n\neq 0 \, .
\eea
On the other hand, if $m=-n$, then
\bea
[L^{\mathrm{st}}_n, L^{\mathrm{st}}_{-n}] &= &  \oint \frac{\di z}{2\pi i} \bigg[   \tfrac{k}{2}(n^3-n) \V^G_{-1} \p\V^G_1    -(n^3-n) i \sqrt{\tfrac{k}{2}} \p Y  - (n^3+n) i \sqrt{\tfrac{k}{2}} \p T  \bigg] (z)  \nonumber \\
&=& 2nL_0 + (n^3-n) \oint \frac{\di z}{2\pi i} \bigg[   \tfrac{k}{2} \V^G_{-1} \p\V^G_1    - i \sqrt{\tfrac{k}{2}} \big(\p Y  + \p T\big)  \bigg] (z) \, .
\eea
Thus, we recover the Virasoro algebra with central term:
\be
[L^{\mathrm{st}}_n, L^{\mathrm{st}}_m] = (n-m)L^{\mathrm{st}}_{n+m} + \tfrac{k\hat{p}}{2} (n^3-n) \delta_{n+m,0}
\ee
where we have defined
\be
\hat{p} \equiv \oint \frac{\di z}{2\pi i} \Big( \p\ln\V^G_1 - i\sqrt{\tfrac{2}{k}}\, \big( \p Y + \p T \big) \Big) \, .
\ee
Thus, the operators
\be
\label{eqn:virasoro-brst-form}
L^{\mathrm{st}}_m = \oint \frac{\di z}{2\pi i} \bigg\{  \Big( m^2 - \tfrac{k}{2} \Big) i\sqrt{\tfrac{2}{k}}\, \p T\, {\cal V}^G_m  - \tfrac{m}{2} : J_G^+ {\cal V}^G_{m-1} : + \tfrac{m}{2} :J_G^- {\cal V}^G_{m+1}: + i m^2 \sqrt{\tfrac{2}{k}}\, \p Y {\cal V}^G_{m} \bigg\} e^{-i\sqrt{\frac{2}{k}} m ( T+Y )}
\ee
form a space-time Virasoro algebra with central charge $c_{\mathrm{st}} = 6k\hat{p}$.

\subsection{Deformed Model}

As we learned in section \ref{sec:bosonicdeformed}, the deformed case is not much different from the undeformed case if we perform a lattice rotation as described in appendix \ref{app:latticerotations}.  In fact, all we have to do is take the undeformed charges (\ref{eqn:virasoro-brst-form}), make the replacement $(T,\varphi)\rightarrow(\tilde{T},\tilde{\varphi})$ (defined in (\ref{newbasis})), and check the invariance under the identification $(\tilde{T},\tilde{\varphi}) \equiv (\tilde{T},\tilde{\varphi})+2\pi R(\tan\alpha,\sec\alpha)$.  The deformed operators will only be invariant under the identification if
\be
\sqrt{\frac{2}{k}} mR\tan\alpha \in \bb{Z} \, ,
\ee
which is the same condition that we've discussed around (\ref{eqn:bosonic-constraint}).  Thus, these operators are only allowed when
\be
\sqrt{\frac{2}{k}} R\tan\alpha \equiv \frac{\mu}{\rho} \in \bb{Q} \, ,  \qquad  m\in\rho\bb{Z} \, ,
\ee
for relatively prime $\mu,\rho\in\bb{Z}$.  Thus, when (\ref{eqn:bosonic-constraint}) holds, we have the operators
\bea
\hat{L}^{\mathrm{st}}_n &=& \frac{k\hat{p}}{4} \big( \rho - \tfrac{1}{\rho} \big) \delta_{n,0}   +  \frac{1}{\rho} \oint \frac{\di z}{2\pi i} \bigg\{  \Big( \rho^2 n^2 - \tfrac{k}{2} \Big) i\sqrt{\tfrac{2}{k}}\, \p \tilde{T}\, {\cal V}^G_{\rho n}  - \tfrac{\rho n}{2} : J_G^+ {\cal V}^G_{\rho n-1} : + \tfrac{\rho n}{2} :J_G^- {\cal V}^G_{\rho n+1}:    \nonumber \\
&& \qquad \qquad \qquad\qquad\qquad\qquad + i \rho^2 n^2 \sqrt{\tfrac{2}{k}}\, \p Y {\cal V}^G_{\rho n} \bigg\} e^{-i\sqrt{\frac{2}{k}} \rho n ( \tilde{T}+Y )} \, .
\eea
When (\ref{eqn:bosonic-constraint}) fails to hold, we still have the operator $\hat{L}_0^{\mathrm{st}}$.

When (\ref{eqn:bosonic-constraint}) holds, we also have a space-time $\widehat{U(1)}$ generator
\begin{eqnarray}
J_{n\rho}^{\mathrm{st}}=\oint \frac{\di z }{2\pi i}\partial\tilde{\varphi}\,\mathcal{V}_{n \rho}e^{-i\sqrt{\frac{2}{k}}n\rho(\tilde{T}+Y)} \, .
\end{eqnarray}
When it fails to hold, we still have $J_0^{\mathrm{st}}$.


\section{Computation of Surface Charges}
\label{app:charges}
This appendix reviews the formalism of \cite{Barnich:2001jy,Barnich:2003xg,Barnich:2007bf} which we use to compute asymptotically conserved charges for our ten-dimensional theory (\ref{10dtheory}) (see also \cite{Compere:2007az} and appendix A of \cite{Compere:2009qm}).
Our $D$-dimensional theory takes the generic form
\begin{equation}
I = \frac{1}{16 \pi G} \int \, \left( R \, \star {\oneone}
- \frac{1}{2}  \star d \chi \wedge d \chi
-  \frac{1}{2}  e^{\alpha .
\chi } \star \mathbf H \wedge \mathbf  H \right)
,\label{gaction}
\end{equation}
where $\chi$ is a scalar field and $\mathbf H$ is a three-form field strength. We will denote the set of fields by $\phi = (g, \mathbf B, \chi)$, where $\mathbf B$ is a two-form potential for $ \mathbf  H$.  Associated to every asymptotic Killing vector $\xi$,\footnote{Asymptotic Killing vectors are defined as diffeomorphisms that satisfy the Killing equations in an asymptotic region and are associated with finite, conserved, and integrable charges.} there is a space-time $D-2$ form
\begin{eqnarray}
\mathbf k_{\xi} [\delta \phi ; \phi ] \label{oneform}
\end{eqnarray}
that is linear in $\delta\phi$ and its derivatives --- it is a one-form in `field space'.  $\mathbf k_{\xi} [\delta \phi ; \phi ]$, which can be constructed by a well-defined algorithm that depends only on the equations of motion, is the basic ingredient in the definition of asymptotically conserved charges \cite{Barnich:2001jy,Barnich:2003xg,Barnich:2007bf} (a similar expression exists for any gauge symmetry parameter of the theory).
It enjoys the following properties:
\begin{itemize}
\item  Given a solution to the equations of motion, $\tilde{\phi}$, and a variation $\delta\phi$ that satisfies the linearized equations of motion around $\phi=\tilde{\phi}$, then for every exact Killing vector $\xi$ of the background $\tilde{\phi}$, there exists a conserved quantity
\beq
\delta Q_{\xi} \equiv \oint_S \mathbf k_{ \xi} [\delta \phi ; \tilde\phi ]  \label{infcharge}
\eeq
that only depend on the homology class of the $(D-2)$-surface $S$.  $\delta Q_\xi$ defines the difference in charge between the backgrounds $\tilde{\phi}$ and $\tilde{\phi}+\delta\phi$ and is unique \cite{Barnich:1994db}.

\item When $\xi$ is an asymptotic Killing vector, the difference in charge between the solutions $\tilde{\phi}$ and $\tilde{\phi}+\delta\phi$ is given by
\beq
\delta Q_{\xi} \equiv \underset{r \rightarrow \infty}{\text{lim}} \oint_{S^r} \mathbf k_{ \xi} [\delta \phi ; \tilde\phi ] \, . \label{infchargeasympt}
\eeq

\item Since $\mathbf k_{ \xi} [\delta \phi ; \tilde\phi ]$ is constructed purely of the equations of motion and solutions $\tilde{\phi}$ and $\tilde{\phi}+\delta\phi$, it does not depend on boundary terms in the action.

\item Since $\mathbf k_{ \xi} [\delta \phi ; \phi ]$ is a linear functional of the equations of motion, it can be expressed as a sum of terms arising from each contribution to the Lagrangian.

\item Given two solutions, $\bar \phi$ and $\tilde\phi$, in the same phase space, for each asymptotic Killing vector $\xi$, the difference in charge between $\bar\phi$ and $\tilde\phi$ is given by
\begin{equation}\label{finitecharge}
Q_{\xi}[\tilde\phi,\bar \phi] \equiv \lim_{r\rightarrow\infty} \oint_{S^r} \int_
\gamma \mathbf{k}_{\xi}[\delta \phi',\phi'] + N_{\xi}[\bar \phi]\, ,
\end{equation}
where $\gamma$ is a path in field space connecting $\bar\phi$ with $\tilde\phi$, $\delta\phi$ and its derivatives are a basis for the line element along $\gamma$, and $N_ {\xi}[\bar \phi]$ is an arbitrary normalization
constant.  Demanding that the charge be independent of the path $\gamma$ implies an integrability condition that restricts the field space of $\phi$ as well as the space of asymptotic Killing vectors.

\end{itemize}
Additional properties of the charge form \eqref{oneform} are discussed in \cite{Barnich:2004uw,Barnich:2006av}.

For the Lagrangian \eqref{gaction}, the contributions to the $(D-2)$-form can be split into four pieces:
\beq
\mathbf k_{ \xi} [\delta \phi ; \phi ] &=&
\mathbf k^{g}_{ \xi}[\delta g;g]  +  e^{\alpha \chi} k^{\mathbf B}_{ \xi}[\delta \phi ; \phi ] +  \mathbf k^{\chi}_{ \xi}
[\delta \phi;\phi]   + \, \mathbf k^{\mathbf B \, suppl}_
{\xi}[\delta \phi;\phi]   \, . \label{k_totg}
\eeq
The gravitational contribution to the charge form is given by \cite{Abbott:1981ff,Barnich:2001jy}
\begin{eqnarray}
\mathbf k^{g}_{ \xi}[\delta g;g] &=& -\delta \mathbf Q^g_{ \xi} -i_{\xi}\mathbf \Theta^g[\delta g] -\mathbf
E^g_\cL[\cL_\xi g, \delta g]\, ,  \label{grav_contrib}
\end{eqnarray}
where
\begin{subequations}
\begin{eqnarray}
\mathbf Q^g_{\xi} &=&  \star \Big(\frac{1}{2} (D_\mu \xi_\nu-D_\nu \xi_\mu)
dx^\mu \wedge dx^\nu \Big)\, ,\label{Komar_term} \\
\mathbf \Theta^g[\delta g]&=&\star \Big(  (D^\sigma \delta
g_{\mu\sigma}-
g^{\alpha\beta} D_\mu \delta g_{\alpha\beta})\,dx^\mu\Big)\, ,\\
\mathbf E^g_\cL[\delta_2 g, \delta_1 g] &=& \star \Big( \frac{1}{2} \delta_1
g_{\mu\alpha} g^{\alpha\beta }\delta_2 g_{\beta\nu} dx^\mu \wedge
dx^\nu \big)\, .
\end{eqnarray}
\end{subequations}
The term \re{Komar_term} is known as the Komar $(D-2)$-form while $E^g_\cL$, which does not appear in the Iyer-Wald
formalism~\cite{Iyer:1994ys}, vanishes for exact Killing vectors but may be relevant for asymptotic symmetries. In \eqref{grav_contrib} above and \eqref{Bcharge} below, $\delta$ is an operator that acts only on the fields $\phi$, not on the asymptotic Killing vectors $\xi$.
The $p$-form contribution to the charge form (here $p=2$) is given by \cite{Compere:2007vx}
\begin{equation}
\mathbf k^{\mathbf B}_{ \xi}[\delta \phi ; \phi ]=-
\delta \mathbf Q^{\mathbf B}_{\xi} + i_\xi \mathbf
\Theta_{\mathbf B}-\mathbf E^{\mathbf B}_\cL[\cL_\xi \mathbf B,\delta \mathbf B] \, , \label{Bcharge}
\end{equation}
where
\begin{eqnarray}
& &\mathbf{Q}^{\mathbf B}_{\xi}  =  i_\xi \mathbf B
\wedge \star \mathbf H \label{QA}\, ,\qquad\qquad\qquad \mathbf \Theta^{\mathbf B} = \delta \mathbf B
\wedge \star \mathbf H\,,\label{ThetaA}\\
& &\mathbf E^{\mathbf B}_\cL[\delta_2 \mathbf B,\delta_1 \mathbf B] =  \star \big(\frac{1}{2}
\frac{1}{(p-1)!}\delta_1
\mathbf B_{\mu\alpha_1\cdots \alpha_{p-1}} \delta_2 \mathbf B_{\nu}^{\;\,\,\alpha_1\cdots \alpha_{p-1}} dx^\mu\wedge
dx^\nu \big)\, .
\end{eqnarray}
Finally, the last two terms are given by
\beq
\mathbf k^{\chi}_{ \xi}[\delta \phi;\phi] &=& i_\xi \big(\star (d
\chi \delta \chi  ) \big) \, , \\ 
\mathbf k^{\mathbf B \, suppl}_{\xi}[\delta \phi; \phi] &= &  \alpha \, \delta \chi \; e^{-
\alpha . \chi} {\cal Q}^{\mathbf B}_{\xi} \, .
\eeq

The next step is the representation of the algebra of asymptotic Killing vectors by the asymptotically conserved charges \re{finitecharge}. For this, we need to define a set of fields  $(\phi, \delta \phi)$ (the \emph{phase space} of the theory) and gauge parameters $\xi$ (the \emph{asymptotic symmetries}) such that the charges $Q_\xi[\phi,\bar \phi]$ are all finite, asymptotically conserved, and integrable for all $\phi$ and $\bar \phi$ in the phase space.
One can then show (modulo a technical assumption) that for any solutions $\phi$ and $\bar\phi$ in the
phase space, and for any asymptotic symmetries $\xi, \xi^\prime, \lambda^
\prime$, the Dirac bracket defined by
\begin{equation}
\left\{ Q_{\xi}[\phi,\bar \phi],
Q_{\xi^\prime}[\phi,\bar \phi] \right\} \equiv
\oint_{S^\infty} \mathbf{k}_{\xi}[\cL_{\xi^\prime}
\phi,;\phi] \label{poissonbracket}
\end{equation}
can be written as
\begin{equation}
\left\{ Q_{\xi}[\phi,\bar \phi],
Q_{\xi^\prime}[\phi,\bar \phi] \right\} =
Q_{[\xi,\xi^\prime]}
[\phi,\bar \phi] -
N_{[\xi,\xi^\prime]}
[\bar \phi] + K_{\xi,\xi^\prime}[\bar \phi]\, ,\label{formula}
\end{equation}
where \begin{eqnarray}
K_{\xi,\xi^\prime}[\bar
\phi] = \oint_{S^\infty} \mathbf{k}_{\xi}[\cL_{\xi^\prime}
\bar\phi;\bar \phi] \label{eq:cc}
\end{eqnarray}
is a central extension that is nontrivial only if it
cannot be reabsorbed into the normalization $N_{[\xi, \xi^\prime]}[\bar \phi]$. An important observation is that the central term can be computed from the data of a background only, independent from the definition of a phase space (and when the phase space is known, the result is independent of the choice of a background in the phase space).

\bibliographystyle{utphys}
\bibliography{master4}

\end{document}